\documentclass[aps,prd,amsmath,floats,floatfix,twocolumn,nofootinbib]{revtex4}
\usepackage{graphicx}
\usepackage{bm}

\newcommand{\twothirds}{{\frac{2}{3}}}
\newcommand{\half}{{\frac{1}{2}}}
\newcommand{\fourth}{{\frac{1}{4}}}
\newcommand{\threefourths}{{\frac{3}{4}}}

\newcommand{\fifth}{{\frac{1}{5}}}
\newcommand{\eighth}{{\frac{1}{8}}}

\newcommand{\fivethirds}{{\frac{5}{3}}}

\newcommand{\kernedcalC}{{\kern 0.1 em{\cal C}}}

\begin{document}

\title{Boundary Conditions for the Einstein Evolution System}

\author{Lawrence E. Kidder${}^1$, Lee Lindblom${}^2$, Mark A. Scheel${}^2$,\\
Luisa T. Buchman${}^3$, and Harald P. Pfeiffer${}^2$}

\affiliation{${}^1$Center for Radiophysics and Space Research, 
Cornell University, Ithaca, New York, 14853}

\affiliation{${}^2$Theoretical Astrophysics 130-33, California Institute of
Technology, Pasadena, CA 91125}

\affiliation{${}^3$Jet Propulsion Laboratory,
4800 Oak Grove Drive, Pasadena, CA 91109}

\date{\today}

\begin{abstract}
New boundary conditions are constructed and tested numerically for a
general first-order form of the Einstein evolution system.  These
conditions prevent constraint violations from entering the
computational domain through timelike boundaries, allow the simulation
of isolated systems by preventing physical gravitational waves from
entering the computational domain, and are designed to be compatible
with the fixed-gauge evolutions used here.  These new boundary
conditions are shown to be effective in limiting the growth of
constraints in 3D non-linear numerical evolutions of dynamical
black-hole spacetimes.
\end{abstract}


\maketitle
%%%%%%%%%%%%%%%%%%%%%%%%%%%%%%%%%%%%%%%%%%%%%%%%%%%%%%%%%%%%%%%%%%%%%%%%%%%%%%
\section{Introduction}
\label{s:Introduction}

The Einstein system can be written as a set of evolution equations
that determine how the dynamical fields change with time, plus
constraint equations that must be satisfied by the physically relevant
field configurations.  The evolution equations ensure that the
constraints will be satisfied within the domain of dependence of the
initial data if they are satisfied initially.  But this does not
guarantee that constraints that are initially small (rather than
precisely zero) will remain small, or that constraint violations will
not enter a domain through its timelike boundaries.  Indeed, the rapid
growth of constraint violations from small (truncation or even
roundoff level) values in the initial data continues to be one of the
major problems for the numerical relativity community.

Constraint violations in the continuum evolution equations (as
distinct from their discrete numerical approximations) have at least two
different causes: {\it a)} bulk constraint violations, in which
existing violations are amplified by the evolution equations, and {\it
b)} boundary violations, in which constraint violations flow into the
domain through timelike boundaries.  A variety of techniques have been
introduced to control bulk constraint violations in numerical
solutions of constrained evolution systems; these include constraint
projection~\cite{Schnetter2003, Anderson2003, Holst2004} (where the
constraint equations are re-solved whenever the constraints become too
large), fully constrained evolution~\cite{piranstark86, AbEv92, ch93,
AbEv93, ACST94, Choptuik2003, Choptuik2003a, Bonazzola2004} (where
some dynamical fields are determined at each time step by solving the
constraint equations rather than their evolution equations), and
dynamical constraint control~\cite{Tiglio2003a, Tiglio2003b,
Lindblom2004} (in which the evolution equations are modified
dynamically in a way that minimizes the growth of the constraints).
Techniques have also been proposed to control boundary constraint
violations; these include the construction of special forms of the
evolution equations that prevent constraint violations from flowing
into the domain~\cite{FriedrichNagy1999, Iriondo2002}, and special
boundary conditions that prevent the influx of constraint violations
in more general forms of the evolution equations~\cite{Stewart1998,
Bardeen2002, Szilagyi2002, Calabrese2002a, Szilagyi2003,
Calabrese2003, Frittelli2003a, Frittelli2003b, Calabrese2003a,
Frittelli2003c, Lindblom2004, Alekseenko2004, Holst2004}.

Recent studies have shown that bulk constraint violations are not
effectively controlled by constraint projection~\cite{Holst2004} or
dynamical constraint control~\cite{Lindblom2004} techniques, unless
the influx of boundary constraint violations is also controlled
separately.  Thus the explicit control of boundary constraint
violations appears to be an essential requirement for effective
constraint control in the Einstein system.  The primary purpose of
this paper is to construct and then test numerically the special
boundary conditions needed to prevent the influx of constraint
violations in one (rather general) first-order form of the Einstein
evolution system: the multi-parameter generalization of the
Frittelli-Reula system~\cite{frittelli_reula96} introduced by Kidder,
Scheel, and Teukolsky (KST)~\cite{Kidder2001}.  These constraint
preserving boundary conditions are constructed here using the strategy
first outlined by Stewart~\cite{Stewart1998} for the Einstein
equations.  The idea is to decompose the constraint fields into
incoming and outgoing parts, based on the characteristic decomposition
of the constraint evolution equations.  The incoming constraint fields
are controlled ({\it e.g.}  set to zero) at each boundary point, and
these conditions then serve as boundary conditions for the principal
dynamical fields of the system.  The constraint preserving boundary
conditions derived here generalize earlier work by being applicable to
generic non-linear field configurations~\cite{Szilagyi2002,
Calabrese2003}, having no symmetry requirements~\cite{Bardeen2002},
and allowing arbitrary values of the gauge fields ({\it i.e.,} the lapse
and shift)~\cite{FriedrichNagy1999, Szilagyi2003}.  These new boundary
conditions are also tested here under more challenging
conditions---non-linear 3D evolutions of dynamical black-hole
spacetimes---than previously considered.

A secondary purpose of this paper is to introduce and test boundary
conditions for the physical (gravitational wave) degrees of freedom of
the Einstein system.  These new physical boundary conditions control
the influx of the radiative part of the Weyl tensor, and generalize
(to the generic 3D case) the boundary conditions of this type proposed
by Bardeen and Buchman~\cite{Bardeen2002}.  Boundary conditions are
also introduced and tested here for those dynamical fields that are
not fixed by the physical or the constraint preserving boundary
conditions.  The boundary conditions for these extra ``gauge'' degrees
of freedom are set in a way that is consistent with the
``fixed-gauge'' ({\it i.e.,} time independent lapse density and
shift) evolutions used in the numerical tests described here.

The remainder of this paper is organized as follows.  The basic KST
form of the Einstein evolution system is reviewed in
Sec.~\ref{s:PrincipalEvolutionSystem}, and additional technical
details (which considerably generalize previously published work) on
the characteristic decomposition of the dynamical fields of this
system are given in Appendix~\ref{s:AppendixA}.  The constraint
evolution equations associated with the KST system are described in
Sec.~\ref{s:ConstraintEvolutionSystem}, and additional details on the
hyperbolicity of the constraint system are given in
Appendix~\ref{s:AppendixB}. The principal new analytical results of
this paper are contained in Sec.~\ref{s:ConstraintPreservingBC}, where
constraint preserving boundary conditions are derived for the KST
system.  Two types of constraint preserving boundary conditions are
presented here: the first, in Sec.~\ref{s:TypeI}, are based on the
characteristic constraint decomposition ideas of
Stewart~\cite{Stewart1998}, while the second, in Sec.~\ref{s:TypeII},
are generalizations of a simpler (but less generally applicable) type
of constraint preserving boundary condition which has been used
effectively for the scalar-wave system~\cite{Holst2004}.  Physical
boundary conditions that control the influx of the radiative part of
the Weyl tensor are presented in Sec.~\ref{s:PhysicalBC}.
Boundary conditions for the remaining ``gauge'' dynamical fields are
given in Sec.~\ref{s:GaugeFixingBC}.  We test the efficacy of these
new boundary conditions by performing numerical evolutions of various
perturbed and unperturbed black-hole spacetimes.  The basic numerical
methods used in these tests are described in
Sec.~\ref{s:NumericalMethods}. The tests themselves are described
in Sec.~\ref{s:NumericalResults}: tests performed on unperturbed black
holes are given in Sec.~\ref{s:Schwarzschild}, tests on perturbed
black holes are presented in Sec.~\ref{sec:pert-schw-black}, and
finally a mild ``angular'' numerical instability that appears in some
of these tests is discussed in Sec.~\ref{sec:angular-stability}.  The major
results of this paper are summarized and outstanding questions
raised by this work are discussed in Sec.~\ref{s:Discussion}.

%%%%%%%%%%%%%%%%%%%%%%%%%%%%%%%%%%%%%%%%%%%%%%%%%%%%%%%%%%%%%%%%%%%%%%%%%%%%%%
\section{Principal Evolution System}
\label{s:PrincipalEvolutionSystem}
The form of the Einstein evolution system used here is a
first-order formulation introduced by Kidder, Scheel, and Teukolsky
(KST)~\cite{Kidder2001}, which generalizes several earlier forms of
the Einstein system~\cite{frittelli_reula96, anderson_york99,
Hern1999}.  This system consists of first-order evolution equations
for the spatial metric $g_{ij}$, the extrinsic curvature $K_{ij}$, and
the spatial derivatives of the metric $D_{kij}=\partial_k
g_{ij}/2$.  The principal (or highest derivative) parts of these
evolution equations are given by
\begin{eqnarray}
&&\!\!\!\!\!
\partial{}_t g{}_{ij}\simeq N^n\partial_n g_{ij},
\label{e:kstggauge}\\
&&\!\!\!\!\!
\partial{}_t K{}_{ij}\simeq N^n\partial_n K_{ij}
-N\Bigl[(1+2\gamma_0)g^{cd}\delta^n{}_{(i}\delta^b{}_{j)}
\nonumber\\
&&\qquad\quad-(1+\gamma_2)g{}^{nd}\delta{}^b{}_{(i}\delta{}^c{}_{j)} 
-(1-\gamma_2) g{}^{bc}\delta{}^n{}_{(i}\delta{}^d{}_{j)} 
\nonumber\\&&\qquad\quad 
+g{}^{nb}\delta{}^c{}_i\delta{}^d{}_j 
+2\gamma_1 g{}^{n[b}g{}^{d]c}g{}_{ij}\Bigr]
\partial{}_n D{}_{bcd},\label{e:kstkgauge}
\end{eqnarray}
\begin{eqnarray}
&&\!\!\!\!\!
\partial{}_t D{}_{kij} \simeq N^n\partial{}_n D{}_{kij}
-N\Bigl[
\delta{}^n{}_k\delta{}^b{}_i\delta{}^c{}_j 
-\half\gamma_3 g{}^{nb}g{}_{k(i}\delta{}^c{}_{j)}\nonumber\\
&&\qquad\quad
-\half\gamma_4 g{}^{nb}g{}_{ij}\delta{}^c{}_{k}
+ \half\gamma_3 g{}^{bc}g{}_{k(i}\delta{}^n{}_{j)}
\nonumber\\
&&\qquad\quad
+ \half\gamma_4 g{}^{bc}g{}_{ij}\delta{}^n{}_{k}\Bigr]\partial{}_n K{}_{bc}
,
\qquad\label{e:kstdgauge}
\end{eqnarray}
where $\simeq$ indicates that terms algebraic in the fields are not
shown explicitly.  This form of the equations assumes that the
lapse density, 
\begin{equation}
Q=\log\left(N g^{-\gamma_0}\right),\label{e:ntilde}
\end{equation}
and the shift $N^i$ are specified {\it a priori} as functions of the
coordinates rather than being evolved as independent dynamical fields.
The parameter\footnote{The
parameters $\gamma_0$ to
$\gamma_4$ used here are related to the parameters of
Ref.~\cite{Kidder2001} by $\gamma_0=\sigma$, $\gamma_1=\gamma$,
$\gamma_2=\zeta$, $\gamma_3=\eta$, and $\gamma_4=\chi$.} 
$\gamma_0$ that appears in these equations is part of
the definition of the lapse density, Eq.~(\ref{e:ntilde}), while
the parameters $\gamma_1$, $\gamma_2$, $\gamma_3$, and $\gamma_4$ were
introduced by adding multiples of the constraints to the standard ADM
form of the evolution equations (see KST~\cite{Kidder2001}).  This
form of the Einstein equations is a quasi-linear first-order system
that can be written more abstractly as
\begin{eqnarray}
\partial_t u^\alpha+A^{k\alpha}{}_{\beta}\partial_k u^\beta &=& F^\alpha,
\label{eq:GeneralFoshSystem}
\end{eqnarray}
where $u^\alpha=\{g_{ij},K_{ij},D_{kij}\}$ is the thirty dimensional
vector of dynamical fields, and the quantities $A^{k\alpha}{}_\beta$
and $F^\alpha$ depend on $u^\alpha$ but not its derivatives
$\partial_k u^\alpha$.  In this paper the Greek indices $\alpha$ and
$\beta$ are used to label elements of this space of
dynamical fields.

Boundary conditions for hyperbolic evolution systems like
Eq.~(\ref{eq:GeneralFoshSystem}) are imposed on the incoming
characteristic fields of the system at each boundary point.  These
characteristic fields are defined as follows.  Given a direction field
$n_k$ ({\it e.g.} the outward directed unit vector normal to the
boundary) we define the left eigenvectors $e^{\hat \alpha}{}_\alpha$
of the characteristic matrix $n_kA^{k\,\alpha}{}_\beta$ by
\begin{eqnarray}
e^{\hat\alpha}{}_\alpha n_k A^{k\,\alpha}{}_\beta=v_{(\hat\alpha)} 
e^{\hat\alpha}{}_\beta,\label{e:eigenvalueeq}
\end{eqnarray}
where $v_{(\hat \alpha)}$ is the eigenvalue (also called the
characteristic speed).  Greek indices with hats, like $\hat\alpha$,
label the various eigenvectors and corresponding eigenvalues of the
system.  There is no summation over $\hat\alpha$ in
Eq.~(\ref{e:eigenvalueeq}).  These eigenvectors form a complete set in
any strongly hyperbolic system of equations, so the matrix
$e^{\hat\alpha}{}_\alpha$ is invertible in this case.  
The projections of the dynamical fields $u^\alpha$ onto these
characteristic eigenvectors are called the characteristic fields
$u^{\hat\alpha}$ of the system:
\begin{eqnarray}
u^{\hat\alpha} = e^{\hat\alpha}{}_{\beta}u^\beta.
\label{e:characteristicparts}
\end{eqnarray}
Boundary conditions must be imposed on any characteristic field having
a negative characteristic speed $v_{(\hat\alpha)}<0$ ({\it i.e.,} an
incoming field) at a particular boundary point.  

The characteristic fields of the KST system are the collection of
fields $u^{\hat\alpha}=$ $\{g_{ij},Z^1$, $Z^2_i$, $Z^3_i$, $Z^4_i$,
$Z^5_{ij}$, $Z^6_{kij}$, $U^{1\pm}$, $U^{2\pm}_i$, $U^{3\pm}$,
$U^{4\pm}_{ij}\}$.  The definitions and explicit expressions for these
fields are given in Appendix~\ref{s:AppendixA}.  The characteristic
fields $U^{1\pm}$, $U^{2\pm}_i$ and $U^{3\pm}$ have characteristic
speeds $\pm v_1$, $\pm v_2$, and $\pm v_3$ respectively relative to
the hypersurface orthogonal observers, where
\begin{eqnarray}
v_1^2&=&2\gamma_0,\label{e:v1}\\
v_2^2&=&\eighth\gamma_3(1-3\gamma_2-4\gamma_0)-\fourth\gamma_4(1+6\gamma_0),
\label{e:v2}\\
v_3^2&=&\half(1+2\gamma_1)(2-\gamma_3+2\gamma_4)-\half \gamma_2\gamma_3.
\label{e:v3}
\end{eqnarray}
The characteristic fields $U^{4\pm}_{ij}$ have speeds $\pm 1$, and
$\{g_{ij}$, $Z^1$, $Z^2_i$, $Z^3_i$, $Z^4_i$, $Z^5_{ij}$,
$Z^6_{kij}\}$ have characteristic speed zero relative to the
hypersurface orthogonal observers.  

The KST evolution system is strongly hyperbolic if and only if the
characteristic fields are linearly independent.  This is the case if
$v_1\neq0$, $v_2\neq 0$, $v_3\neq 0$, and $v_1\neq v_3$.  The system
is also strongly hyperbolic for $v_1=v_3\neq 0$ and $1+3v_1^2=4v^2_2$.
In the strongly hyperbolic case the fundamental characteristic fields
have thirty independent components: the spatial metric $g_{ij}$ (with
six independent components), five scalars $\{Z^1,U^{1\pm},U^{3\pm}\}$,
five transverse (to $n^k$) vectors $\{Z^2_i,Z^3_i,Z^4_i,U^{2\pm}_i\}$
(which have a total of ten independent components), three
transverse-traceless second-rank tensors $\{Z^5_{ij}$,
$U^{4\pm}_{ij}\}$ (which have a total of seven independent components,
since $U^{4\pm}_{ij}$ is symmetric while $Z^5_{ij}$ is not), and one
transverse-traceless third-rank tensor $Z^6_{kij}$ (which has a total
of two independent components).  The KST system also
admits a positive-definite symmetrizer matrix $S_{\alpha\beta}$ (and
so is symmetric hyperbolic) under fairly weak conditions on the KST
parameters, which are discussed in detail in Appendix B of
Ref.~\cite{Lindblom2003}.

The characteristic speeds relative to the coordinate
frame, {\it i.e.,} the eigenvalues $v_{(\hat\alpha)}$ of
Eq.~(\ref{e:eigenvalueeq}), are $\pm Nv_1 - n_kN^k$, $\pm
Nv_2-n_kN^k$, $\pm Nv_3-n_kN^k$ and $\pm N-n_kN^k$ for the fields
$U^{1\pm}$, $U^{2\pm}_i$, $U^{3\pm}$, and $U^{4\pm}_{ij}$
respectively.  And the speeds are $-n_kN^k$ for the ``zero-speed'' fields
$\{g_{ij}$, $Z^1$, $Z^2_i$, $Z^3_i$, $Z^4_i$, $Z^5_{ij}$,
$Z^6_{kij}\}$.  Boundary conditions must be supplied for each incoming
characteristic field, {\it i.e.,} for each field whose coordinate
characteristic speed is negative at a particular boundary point.  The
fields $U^{1-}$, $U^{2-}_i$, $U^{3-}$, and $U^{4-}_{ij}$ will be
ingoing at most timelike boundaries; the zero-speed fields
$\{g_{ij}$, $Z^1$, $Z^2_i$, $Z^3_i$, $Z^4_i$, $Z^5_{ij}$,
$Z^6_{kij}\}$ will be ingoing at any boundary where the shift
$N^i$ is directed out of the computational domain, {\it i.e.,}
whenever $n_kN^k>0$.  Boundary conditions must be formulated therefore
for each of the fields $\{g_{ij}$, $Z^1$, $Z^2_i$, $Z^3_i$, $Z^4_i$,
$Z^5_{ij}$, $Z^6_{kij}$, $U^{1-}$, $U^{2-}_i$, $U^{3-}$, $U^{4-}_{ij}\}$.
%
%%%%%%%%%%%%%%%%%%%%%%%%%%%%%%%%%%%%%%%%%%%%%%%%%%%%%%%%%%%%%%%%%%%%%%%%%%%%%%%
\section{Constraint Evolution System}
\label{s:ConstraintEvolutionSystem}
Boundary conditions capable of preventing the influx of constraint
violations cannot be formulated without a complete understanding of
how the constraints propagate.  The constraints associated with the
(vacuum) Einstein evolution system are
\begin{eqnarray}
{\cal C}        &=& \half[{}^{(3)}R - K_{ij}K^{ij} + K^2],\label{e:hamcdef}\\
{\cal C}_i      &=& \nabla_j K^j{}_i - \nabla_i K,\\
{\cal C}_{kij}  &=& \partial_k g_{ij}-2D_{kij},\label{e:c3def}\\ 
{\cal C}_{klij} &=& 2\partial_{[k}D_{l]ij}. \label{e:c4def}
\end{eqnarray}
Here ${\cal C}$ is the Hamiltonian constraint, ${\cal C}_i$ is the
momentum constraint, while ${\cal C}_{kij}$ and ${\cal C}_{klij}$ are
auxiliary constraints associated with the introduction of the
dynamical field $D_{kij}$ needed to make the KST system first order.
All of these constraint fields are zero for the physical solutions of
the (vacuum) Einstein evolution system.  Note that ${\cal C}_{kij}$
and ${\cal C}_{klij}$ are not completely independent: there is a
second-class constraint ${\cal C}_{klij} =\partial_{[l}{\cal
C}_{k]ij}$. However, ${\cal C}_{kij}$ and ${\cal C}_{klij}$ must both
be retained and treated as independent in order to write the evolution
of the constraints as a closed first-order hyperbolic system, which is
the goal of this section.

The evolution of the constraints is completely determined by the
evolution of the dynamical fields of the principal evolution system.
Using the KST evolution equations,
Eqs.~(\ref{e:kstggauge})--(\ref{e:kstdgauge}), and the definitions of
the constraints, Eqs.~(\ref{e:hamcdef})--(\ref{e:c4def}), it is
straightforward to show that the principal parts of the constraint
evolution equations are
\begin{eqnarray}
\partial_t{\cal C} &\simeq & N^k\partial_k {\cal C}
-\half(2-\gamma_3+2\gamma_4) N g^{ij}\partial_i{\cal C}_j,
\label{e:hamdot}\\
\partial_t{\cal C}_i &\simeq & N^k\partial_k{\cal C}_i
- (1+2\gamma_1) N \partial_i {\cal C}\nonumber\\
&&+\half N g^{kl}g^{ab}\Bigl[(1-\gamma_2)\partial_k{\cal C}_{labi}
+(1+\gamma_2)\partial_k{\cal C}_{ailb}\nonumber\\
&&\quad\qquad\qquad-(1+2\gamma_0)\partial_k{\cal C}_{liab}\Bigr],\\
\partial_t{\cal C}_{kij}&\simeq & N^k\partial_k{\cal C}_{kij},
\label{e:c3dot}\\
\partial_t{\cal C}_{klij} &\simeq & N^m\partial_m{\cal C}_{klij}
+\half \gamma_3 N \bigl(g_{j[l} \partial_{k]}{\cal C}_i
+g_{i[l} \partial_{k]}{\cal C}_j\bigr)\nonumber\\
&&+\gamma_4 N g_{ij}\partial_{[k}{\cal C}_{l]}.\label{e:c4dot}
\end{eqnarray}
More abstractly, the constraints satisfy a quasi-linear evolution
system of the form
\begin{eqnarray}
\partial_t c^A + A^{kA}{}_B \partial_k c^B = F^A{}_B c^B,
\label{e:constevsystem}
\end{eqnarray}
where $c^A=\{{\cal C}$, ${\cal C}_i$, ${\cal C}_{kij}$, ${\cal
C}_{klij}\}$ are the constraints defined in
Eqs.~(\ref{e:hamcdef})--(\ref{e:c4def}), $A^{kA}{}_B$ depends on the
dynamical fields $u^\alpha$, and $F^A{}_B$ depends on $u^\alpha$ and
$\partial_k u^\alpha$.  We use upper case Latin indices such as
$\scriptstyle A$ and $\scriptstyle B$ to label the constraint fields.
If the constraint evolution system is hyperbolic, then
Eq.~(\ref{e:constevsystem}) guarantees that the constraints will
vanish everywhere if they are zero at some initial time, and if
boundary conditions are chosen to prevent the influx of constraints
through timelike boundaries.

In order to determine whether this constraint evolution system is
hyperbolic, we evaluate the characteristic fields associated with
Eq.~(\ref{e:constevsystem}).  It is straightforward to show that the
fields $c^{\hat A}=\{{\cal C}_{kij}$, $Z^7$, $Z^8_i$, $Z^9_i$,
$Z^{10}_{ij}$, $Z^{12}_{ijkl}$, $U^{5\pm}_i$, $U^{6\pm}\}$ are
characteristic constraint fields, where the individual components of
$c^{\hat A}$ are defined by
\begin{eqnarray}
Z^7&=& \gamma_3 {\cal C} -(2-\gamma_3+2\gamma_4)n^k n^l {\cal
  C}^2_{kl},\label{e:z7def}\\
Z^8_i&=&n^kP^l{}_i\bigl[\gamma_4 {\cal C}^1_{kl}
-(\gamma_3+3\gamma_4)n^an^b{\cal C}_{klab}\bigr],\label{e:z8def}\\
Z^9_i&=&n^kP^l{}_i\bigl(3 {\cal C}^1_{kl}+2 {\cal C}^2_{kl}
-7n^an^b{\cal C}_{klab}\bigr),\label{e:z9def}\\
Z^{10}_{ij}&=&P^k{}_iP^l{}_j{\cal C}^1_{kl},\\
Z^{11}_{ij}&=&\big(P^k{}_iP^l{}_j-\half P_{ij}P^{kl}\big)
{\cal C}^2_{kl},\label{e:z11def}\\
Z^{12}_{ijkl}&=&{\cal C}_{ijkl}
-{\scriptscriptstyle\frac{3}{5}}{\cal C}^1_{ij}g_{kl}
-{\scriptscriptstyle\frac{2}{5}}
\bigl({\cal C}^1_{k[i}g{}_{j]l}+{\cal C}^1_{l[i}g{}_{j]k}\bigr)\nonumber\\
&&-{\scriptscriptstyle\frac{4}{5}}{\cal C}^2_{[ij]}g_{kl}
-{\scriptscriptstyle\frac{4}{15}}
\bigl({\cal C}^2_{k[i}g{}_{j]l}+{\cal C}^2_{l[i}g{}_{j]k}\bigr)\nonumber\\
&&-{\scriptscriptstyle\frac{14}{15}}
\bigl(g{}_{k[i}{\cal C}^2_{j]l}+g{}_{l[i}{\cal C}^2_{j]k}\bigr),
\label{e:z12def}\\
U^{5\pm}_i&=& \pm n^kP^l{}_i\bigl[(1+2\gamma_0){\cal C}^1_{kl}+
2{\cal C}^2_{[kl]}-2\gamma_2{\cal C}^2_{(kl)}\bigr]\nonumber\\
&&+ 2 v_2 P^l{}_i {\cal C}_l,\label{e:u5def}\\
U^{6\pm}&=&(1+2\gamma_1){\cal C}\pm v_3 n^k{\cal C}_k
-\gamma_2 n^kn^l{\cal C}^2_{kl}.\label{e:u6def}
\end{eqnarray}
The $v_2$ and $v_3$ that appear in
Eqs.~(\ref{e:z7def})--(\ref{e:u6def}) are given by Eqs.~(\ref{e:v2})
and (\ref{e:v3}), and ${\cal C}^1_{ij}$ and ${\cal C}^2_{ij}$ are
defined by
\begin{eqnarray}
{\cal C}^1_{ij}&=&g^{kl}{\cal C}_{ijkl},\\
{\cal C}^2_{ij}&=&g^{kl}{\cal C}_{kijl}.
\end{eqnarray}

The characteristic constraint fields $U^{5\pm}_i$ and $U^{6\pm}$ have
characteristic speeds $\pm v_2$ and $\pm v_3$ respectively, while the
characteristic constraint fields $\{{\cal C}_{kij}$, $Z^7$, $Z^8_i$,
$Z^9_i$, $Z^{10}_{ij}$, $Z^{11}_{ij}$, $Z^{12}_{ijkl}\}$ have
characteristic speed zero relative to the hypersurface orthogonal
observers.  The characteristic constraint fields have forty
independent components: ${\cal C}_{kij}$ (which has
eighteen independent components), three scalars $\{Z^7,U^{6\pm}\}$,
four transverse (to $n^k$) vectors $\{Z^8_i,Z^9_i,U^{5\pm}_i\}$ (which
have a total of eight independent components), one antisymmetric
transverse second-rank tensor $Z^{10}_{ij}$ (which has one independent
component), one transverse traceless second-rank tensor $Z^{11}_{ij}$
(which has three independent components), and one totally trace-free
fourth-rank tensor $Z^{12}_{ijkl}$ (which is antisymmetric in its
first two indices, symmetric in its last two indices, and so has seven
independent components).  These characteristic constraint fields are
linearly independent so long as $v_2\neq0$ and $v_3\neq0$.  Thus the
characteristic constraint evolution equations are strongly hyperbolic
whenever the principal evolution system is strongly hyperbolic.  In
Appendix~\ref{s:AppendixB} we summarize the conditions under which the
constraint evolution system is also symmetric hyperbolic.
%
%%%%%%%%%%%%%%%%%%%%%%%%%%%%%%%%%%%%%%%%%%%%%%%%%%%%%%%%%%%%%%%%%%%%%%%%%%%%%%%
\section{Constraint Preserving Boundary Conditions}
\label{s:ConstraintPreservingBC}
The constraint characteristic fields $U^{5-}_i$ and $U^{6-}$ are
incoming fields at most timelike boundaries, and the fields $\{{\cal
C}_{kij}$, $Z^7$, $Z^8_i$, $Z^9_i$, $Z^{10}_{ij}$, $Z^{11}_{ij}$,
$Z^{12}_{ijkl}\}$ are incoming at any boundary where $n_kN^k>0$.
Therefore we must choose boundary conditions on the incoming
characteristic fields of the principal evolution system,
$u^{\hat\alpha}$, in a way that controls these incoming constraint
characteristic fields, $c^{\hat A}$.  We use two different approaches
to accomplish this for the KST system.  The first approach (introduced
by Stewart~\cite{Stewart1998} and developed by Calabrese, {\it et
al.}~\cite{Calabrese2003}) re-expresses the incoming characteristic
constraint fields, $c^{\hat A}$, in terms of the principal
characteristic fields, $u^{\hat\alpha}$.  This results in a set of
Neumann-like boundary conditions on certain incoming characteristic
fields $u^{\hat\alpha}$.  These {\em Type I} boundary conditions are
discussed in more detail in Sec.~\ref{s:TypeI}.  The second (less
general) approach (introduced by Holst, {\it et al.}~\cite{Holst2004})
uses a more direct Dirichlet-like boundary condition for certain
incoming characteristic fields $u^{\hat\alpha}$.  These {\em Type II}
boundary conditions are discussed in more detail in
Sec.~\ref{s:TypeII}.

We note that it is not possible to use these methods to derive
boundary condition for all of the characteristic fields (of the
principal system) that need
them.  In particular it is not possible to obtain boundary conditions
for the fields $Z^4_i$, $U^{1-}$, and $U^{4-}_{ij}$ in this way.  The
boundary conditions for these fields are determined by physical and
gauge considerations, which are discussed in detail in
Secs.~\ref{s:PhysicalBC} and \ref{s:GaugeFixingBC}.
%
%%%%%%%%%%%%%%%%%%%%%%%%%%%%%%%%%%%%%%%%%%%%%%%%%%%%%%%%%%%%%%%%%%%%%%%%%%%%%
\subsection{Type I Boundary Conditions}
\label{s:TypeI}
The incoming constraint characteristic fields ${\cal C}_{kij}$, $Z^7$,
$Z^8_i$, $Z^9_i$, $Z^{11}_{ij}$, $Z^{12}_{ijkl}$, $U^{5-}_i$, and
$U^{6-}$ all depend on normal ({\it i.e.,} perpendicular to the
boundary) derivatives of the principal characteristic fields $u^{\hat
\alpha}$. A straightforward but lengthy calculation shows that 
these characteristic constraint fields can be expressed as follows:
\begin{eqnarray}
&&\!\!\!\!\!\!\!n^k{\cal C}_{kij}= d_\perp g_{ij} -2 n^k D_{kij},
\label{e:gijcond}\\
&&\!\!\!\!\!\!\!Z^7=-d_\perp Z^1 
-(2-\gamma_3+2\gamma_4)P^{ab}n^cn^d\partial_aD_{cdb}\nonumber\\
&&+\half\gamma_3\Bigl(8D_{[ab]}{}^bD_c{}^{ac}-D_{ab}{}^bD^{ac}{}_c
-K_{ab}K^{ab}+K^2\Bigr)\nonumber\\
&&-2\gamma_3P^{ab}g^{cd}\partial_a D_{[bc]d}
+\half\gamma_3D_{abc}\bigl(3D^{abc}-2D^{cab}\bigr),
\label{e:z7cond}\\
&&\!\!\!\!\!\!\!Z^8_i=d_\perp Z^2_i
-\bigr[\gamma_4g^{ab}-(\gamma_3+3\gamma_4)n^an^b\bigr] 
n^cP^d{}_i\partial_d D_{cab},\\
&&\!\!\!\!\!\!\!Z^9_i=d_\perp Z^3_i
-\bigl[(3g^{ab}-7n^an^b)P^d{}_i-2P^a{}_iP^{bd}\bigr]
n^c\partial_d D_{cab},\nonumber\\
&&\\
&&\!\!\!\!\!\!\!Z^{11}_{ij}= d_\perp Z^5_{ij}
\nonumber\\
&&-\bigl(P^{a}{}_i P^{b}{}_j-\half P_{ij} P^{ab}\bigr)
\bigl(g^{cd}\partial_{a} D_{cbd}-P^{cd}\partial_{c} D_{abd}\bigr),
\label{e:z11cond}
\end{eqnarray}
\begin{eqnarray}
&&\!\!\!\!\!\!\!n^d P^{cab}_{kij}Z^{12}_{dcab}= d_\perp Z^6_{kij}
-n^cP^{dab}_{kij}\partial_d D_{cab},\label{e:z12cond}\\
&&\!\!\!\!\!\!\!U^{5-}_i=-d_\perp U^{2-}_i+2v_2(P^{jk}P^l{}_i-g^{jl}P^k{}_i)
\partial_k K_{jl}
\nonumber\\
&&+2v_2\bigl(g^{aj}g^{bk}P^c{}_i+g^{cj}g^{ab}P^k{}_i
-2g^{ca}g^{bj}P^k{}_i\bigr)D_{cab}K_{jk}\nonumber\\
&&-\bigl[(1+\gamma_2)n^m g^{ln}
-(1+2\gamma_0)n^l g^{mn}\bigr]P^k{}_i\partial_k
D_{lmn}\nonumber\\
&&-\bigl[(1-\gamma_2)n^lP^m{}_i
-(1+\gamma_2)n^mP^l{}_i\bigr]P^{kn}\partial_k D_{lmn},
\label{e:u5minuscond}\\
&&\!\!\!\!\!\!\!U^{6-}= d_\perp U^{3-}-v_3 n^l P^{jk}\partial_k K_{jl}
-\gamma_2 P^{kn} n^l n^m\partial_k D_{lmn}\nonumber\\
&&+(1+2\gamma_1)\bigl(2g^{m[l}P^{n]k}\partial_k D_{lmn}
+g^{i[j}g^{a]b}K_{ij}K_{ab}\bigr)\nonumber\\
&&-v_3\bigl(g^{aj}g^{bk}n^c+g^{cj}g^{ab}n^k
-2g^{ca}g^{bj}n^k\bigr)D_{cab}K_{jk}\nonumber\\
&&-\half(1+2\gamma_1)\Bigl(g^{kc}g^{ij}g^{ab}+2g^{ka}g^{ib}g^{jc}
+8g^{k[i}g^{a]j}g^{cb}
\nonumber\\
&&\qquad\qquad\qquad-3g^{kc}g^{ia}g^{jb}\Bigr)D_{kij}D_{cab},
\label{e:u6minuscond}
\end{eqnarray}
where $P^{cab}_{kij}$ is defined by Eq.~(\ref{e:projectiontensor}),
and the quantities $d_\perp g_{ij}$, $d_\perp Z^1$, $d_\perp Z^2_i$,
$d_\perp Z^3_i$, $d_\perp Z^5_{ij}$, $d_\perp Z^6_{kij}$, $d_\perp
U^{2-}_i$ and $d_\perp U^{3-}$ are components of $d_\perp
u^{\hat\alpha}$, the characteristic projections of the normal
derivatives of the dynamical fields.  These projected normal
derivatives are defined by
\begin{eqnarray}
d_\perp u^{\hat\alpha} \equiv e^{\hat\alpha}{}_\beta n^k \partial_k u^\beta.
\label{eq:dperp}
\end{eqnarray}
We note that Eqs.~(\ref{e:gijcond})--(\ref{e:u6minuscond}) express
these incoming characteristic constraint fields, $c^{\hat A}$,
in terms of the normal derivative of one of the principal 
characteristic fields,
$u^{\hat\alpha}$, plus terms that depend on derivatives 
tangent to the boundary and terms that are algebraic in the fields.

The simplest Type I constraint preserving boundary conditions are
obtained by setting the incoming components of the characteristic
constraint fields to zero.  Using
Eqs.~(\ref{e:gijcond})--(\ref{e:u6minuscond}), these expressions
become boundary conditions for the normal derivatives of the incoming
dynamical fields $d_\perp g_{ij}$, $d_\perp Z^1$, $d_\perp Z^2_i$,
$d_\perp Z^3_i$, $d_\perp Z^5_{ij}$, $d_\perp Z^6_{kij}$, $d_\perp
U^{2-}_i$ and $d_\perp U^{3-}$:
\begin{eqnarray}
&&\!\!\!\!\!\!\!d_\perp g_{ij} = 2 n^k D_{kij},\label{e:gijbc}\\
&&\!\!\!\!\!\!\!d_\perp Z^1= 
-(2-\gamma_3+2\gamma_4)P^{ab}n^cn^d\partial_aD_{cdb}\nonumber\\
&&+\half\gamma_3\Bigl(8D_{[ab]}{}^bD_c{}^{ac}-D_{ab}{}^bD^{ac}{}_c
-K_{ab}K^{ab}+K^2\Bigr)\nonumber\\
&&-2\gamma_3P^{ab}g^{cd}\partial_a D_{[bc]d}
+\half\gamma_3D_{abc}\bigl(3D^{abc}-2D^{cab}\bigr),
\label{e:z1cond}\\
&&\!\!\!\!\!\!\!d_\perp Z^2_i=
\bigr[\gamma_4g^{ab}-(\gamma_3+3\gamma_4)n^an^b\bigr] 
n^cP^d{}_i\partial_d D_{cab},\label{e:z2cond}\\
&&\!\!\!\!\!\!\!d_\perp Z^3_i=
\bigl[(3g^{ab}-7n^an^b)P^d{}_i-2P^a{}_iP^{bd}\bigr]
n^c\partial_d D_{cab},\label{e:z3cond}\\
&&\!\!\!\!\!\!\!d_\perp Z^5_{ij}=
\bigl(P^{a}{}_i P^{b}{}_j-\half P_{ij} P^{ab}\bigr)
\bigl(g^{cd}\partial_{a} D_{cbd}-P^{cd}\partial_{c} D_{abd}\bigr),\nonumber\\
&&\\
&&\!\!\!\!\!\!\!d_\perp Z^6_{kij}=
n^cP^{kab}_{kij}\partial_d D_{cab},\label{e:z6cond}\\
&&\!\!\!\!\!\!\!d_\perp U^{2-}_i=2v_2(P^{jk}P^l{}_i-g^{jl}P^k{}_i)
\partial_k K_{jl}
\nonumber\\
&&+2v_2\bigl(g^{aj}g^{bk}P^c{}_i+g^{cj}g^{ab}P^k{}_i
-2g^{ca}g^{bj}P^k{}_i\bigr)D_{cab}K_{jk}\nonumber\\
&&-\bigl[(1+\gamma_2)n^m g^{ln}
-(1+2\gamma_0)n^l g^{mn}\bigr]P^k{}_i\partial_k
D_{lmn}\nonumber\\
&&-\bigl[(1-\gamma_2)n^lP^m{}_i
-(1+\gamma_2)n^mP^l{}_i\bigr]P^{kn}\partial_k D_{lmn},
\label{e:u2minusbc}
\end{eqnarray}
\begin{eqnarray}
&&\!\!\!\!\!\!\!d_\perp U^{3-}=v_3 n^l P^{jk}\partial_k K_{jl}
+\gamma_2 P^{kn} n^l n^m\partial_k D_{lmn}\nonumber\\
&&-(1+2\gamma_1)\bigl(2g^{m[l}P^{n]k}\partial_k D_{lmn}
+g^{i[j}g^{a]b}K_{ij}K_{ab}\bigr)\nonumber\\
&&+v_3\bigl(g^{aj}g^{bk}n^c+g^{cj}g^{ab}n^k
-2g^{ca}g^{bj}n^k\bigr)D_{cab}K_{jk}\nonumber\\
&&+\half(1+2\gamma_1)\Bigl(g^{kc}g^{ij}g^{ab}+2g^{ka}g^{ib}g^{jc}
+8g^{k[i}g^{a]j}g^{cb}
\nonumber\\
&&\qquad\qquad\qquad-3g^{kc}g^{ia}g^{jb}\Bigr)D_{kij}D_{cab}.
\label{e:u6minusbc}
\end{eqnarray}
More general Type I boundary conditions of this type can also be
obtained by setting the incoming constraint characteristic fields to
multiples of the corresponding outgoing fields: {\it i.e.,} by setting
$U^{5-}_i=\mu_5 U^{5+}_i$ and $U^{6-}=\mu_6 U^{6+}$.  These boundary
conditions can therefore be generalized by adding the terms $\mu_5
U^{5+}_i$ and $- \mu_6 U^{6+}$ to the rights sides of
Eqs.~(\ref{e:u2minusbc}) and (\ref{e:u6minusbc}) respectively.

We point out that the combination of constraints,
\begin{eqnarray}
Z^9_i -15P^a{}_iN^bn^cn^d Z^{12}_{abcd}=16 P^{a[c}P^{d]}{}_in^b
\partial_d D_{cab},
\end{eqnarray}
involves no normal derivatives of $D_{cab}$.  Thus, arbitrary
multiples of this combination of constraints could be added to the
Type I boundary conditions for $Z^2_i$, $Z^3_i$ and $U^{2-}_i$ in
Eqs.~(\ref{e:z2cond}), (\ref{e:z3cond}) and (\ref{e:u2minusbc})
without changing the basic structure of these conditions at all.  We
find that the addition of these terms does change the stability of
numerical evolutions.  But at present we have not found a systematic
way to optimize the addition of these extra constraint fields (short
of brute force numerical testing), so the numerical evolutions
presented here use the simplest analytical forms given in
Eqs.~(\ref{e:z2cond}), (\ref{e:z3cond}) and (\ref{e:u2minusbc}).

We find it convenient to impose these Type I constraint preserving
boundary conditions via the Bj{\o}rhus method~\cite{Bjorhus1995} (which
we commonly use in our numerical code) in the following way.  The
characteristic projection of the principal evolution system can be
written as
\begin{eqnarray}
d_{t} u^{\hat\alpha}+v_{(\hat\alpha)}d_\perp u^{\hat\alpha}=
e^{\hat\alpha}{}_\beta \bigl(-A^{k\beta}{}_\alpha P^i{}_k \partial_i u^\alpha
+F^\beta\bigr),\label{e:projsystem}
\end{eqnarray}
where $d_{t} u^{\hat\alpha}$ represents the characteristic projection of
the time derivative of the dynamical field:
\begin{eqnarray}
d_t u^{\hat\alpha} \equiv e^{\hat\alpha}{}_\beta \partial_t u^\beta.
\label{eq:dtudefinition}
\end{eqnarray}
The terms on the right side of Eq.~(\ref{e:projsystem}) depend on the
boundary values of the dynamical fields and the derivatives of these
fields tangent to the boundary surface.  We leave these terms unchanged.
However, we replace the term
$d_\perp u^{\hat\alpha}$ on the left side of
Eq.~(\ref{e:projsystem}) by its desired value $d_\perp
u^{\hat\alpha}_{\hbox{bc}}$.  For the fields 
$g_{ij}$, $Z^1$, $Z^2_i$, $Z^3_i$,
$Z^5_{ij}$, $U^{2-}_i$ and $U^{3-}$, $d_\perp
u^{\hat\alpha}_{\hbox{bc}}$ is given by
Eqs.~(\ref{e:gijbc})--(\ref{e:u6minusbc}).

This method of imposing these boundary conditions can be implemented
numerically in a simple and elegant way.  Let us define the quantity
\begin{eqnarray}
D_t u^{\hat\alpha} \equiv e^{\hat\alpha}{}_\beta\bigl(-A^{k\beta}{}_\alpha
\partial_k u^\alpha + F^\beta\bigr),\label{e:Dtuhatdef}
\end{eqnarray}
which is just the characteristic projection of the right side of the
full evolution equation, without any modifications for boundary
conditions.  At the boundary, Eq.~(\ref{e:projsystem}) then takes
the form
\begin{eqnarray}
d_{t} u^{\hat\alpha} = D_t u^{\hat\alpha} 
                     + v_{(\hat\alpha)} 
		     \bigl(  d_\perp u^{\hat\alpha}
		           - d_\perp u^{\hat\alpha}_{\hbox{bc}} \bigr).
\label{e:simpletype1bc}
\end{eqnarray}
In this expression, $d_\perp u^{\hat\alpha}_{\hbox{bc}}$ represents
the desired values of the tangential derivatives at the boundary, and
$d_\perp u^{\hat\alpha}$ denotes the actual values
of the tangential derivatives as defined in Eq.~(\ref{eq:dperp}). When
$d_\perp u^{\hat\alpha}$ is expressed in terms of the constraints
on the boundary using
Eqs.~(\ref{e:gijcond})--(\ref{e:u6minuscond}), and when
$d_\perp u^{\hat\alpha}_{\hbox{bc}}$ is determined from 
Eqs.~(\ref{e:gijbc})--(\ref{e:u6minusbc}),
then the term $d_\perp u^{\hat\alpha}-d_\perp u^{\hat\alpha}_{\hbox{bc}}$
simplifies considerably.
Thus in particular, the boundary
conditions for $g_{ij}$, $Z^1$, $Z^2_i$, $Z^3_i$, $Z^5_{ij}$,
$Z^6_{kij}$, $U^{2-}_i$ and $U^{3-}$ can be written as
\begin{eqnarray}
d_t g_{ij} &=& D_t g_{ij} - n_k N^k n^l{\cal C}_{lij},\label{e:gdotbc}\\
d_t Z^1 &=& D_t Z^1 + n_k N^k Z^7,\label{e:z1bc}\\
d_t Z^2_i &=& D_t Z^2_i - n_k N^k Z^8_i,\label{e:z2bc}\\
d_t Z^3_i &=& D_t Z^3_i - n_k N^k Z^9_i,\label{e:z3mbc}\\
d_t Z^5_{ij} &=& D_t Z^5_{ij} - n_k N^k Z^{11}_{ij},\label{e:z11mbc}\\
d_t Z^6_{kij} &=& D_t Z^6_{kij} - n_m N^m n^dP^{cab}_{kij}{Z}^{12}_{dcab},
\label{e:z12mbc}\\
d_t U^{2-}_i &=& D_t U^{2-}_i + (N v_2 + n_k N^k) 
(U^{5-}_i-\mu_5 U^{5+}_i),\qquad\label{e:u2mbc}\\
d_t U^{3-} &=& D_t U^{3-} - (N v_3 + n_k N^k) 
(U^{6-}-\mu_6U^{6+}).\label{e:u3mbc}
\end{eqnarray}
The $D_t u^{\hat \alpha}$ that appear in these expressions are to be
evaluated using Eq.~(\ref{e:Dtuhatdef}), while the constraint fields
${\cal C}_{kij}$, $Z^7$, $Z^8_i$, $Z^9_i$, $Z^{11}_{ij}$,
$Z^{12}_{klij}$, $U^{5-}_i$ and $U^{6-}$ are to be evaluated
numerically using their definitions in Eqs.~(\ref{e:c3def}),
(\ref{e:z7def}), (\ref{e:z8def}), (\ref{e:z9def}), (\ref{e:z11def}),
(\ref{e:z12def}), (\ref{e:u5def}) and (\ref{e:u6def}) respectively.

%%%%%%%%%%%%%%%%%%%%%%%%%%%%%%%%%%%%%%%%%%%%%%%%%%%%%%%%%%%%%%%%%%%%%%%%%%%%%
\subsection{Type II Boundary Conditions}
\label{s:TypeII}

A second type of constraint preserving boundary condition can be
imposed on those characteristic fields that represent various
projections of the field $P^k{}_lD_{kij}$.  The influx of constraint
violations that could cause $P^k{}_lD_{kij}$ to differ from 
$P^k{}_l\partial_kg_{ij}/2$ can be prevented by enforcing the equality of
these fields on the relevant boundaries.  This is a Dirichlet-like
boundary condition that sets $P^k{}_lD_{kij}=P^k{}_l\partial_kg_{ij}/2$.
  This type of constraint preserving boundary
condition has been used successfully in the scalar wave system by
Holst, {\it et al.\/}~\cite{Holst2004}.  The characteristic fields
$Z^2_i$, $Z^3_i$, $Z^5_{ij}$, and $Z^6_{kij}$ are composed entirely of
various components of $P^k{}_lD_{kij}$, which represent the
derivatives of $g_{ij}$ that are tangent to the boundary.  These Type
II boundary conditions can be imposed using the Bj{\o}rhus method
by setting
\begin{eqnarray}
d_t Z^2_i&=&\half\bigl[\gamma_4 P^{ab}-(\gamma_3+2\gamma_4)
n^an^b\bigr]P^c{}_i\partial_c\partial_t g_{ab},\label{e:z2bcII}\\
d_t Z^3_i&=&\half\bigl[(3 P^{ab}-4n^an^b)P^c{}_i-2P^a{}_iP^{bc}
\bigr]\partial_c\partial_t g_{ab},\quad\,\,\label{e:z3bcII}\\
d_t Z^5_{ij}&=&
\fourth\bigl(2P^c{}_iP^a{}_j-P_{ij}P^{ca}\bigr)n^b
\partial_c\partial_t g_{ab},\label{e:z5bcII}\\
d_tZ^6_{kij}&=&
\half P^{cab}_{kij}\partial_c \partial_t g_{ab},
\label{e:z6bcII}
\end{eqnarray}
at any boundary where these fields are incoming.  The quantity
$\partial_c\partial_t g_{ab}$ that appears on the right sides of
Eqs.~(\ref{e:z2bcII})--(\ref{e:z6bcII}) is to be evaluated by taking
the indicated tangential spatial derivatives of the right side of
Eq.~(\ref{e:gdotbc}).  These expressions are obtained using the
expressions for $Z^2_i$, $Z^3_i$, $Z^5_{ij}$, and $Z^6_{kij}$ from
Eqs.~(\ref{e:z2}), (\ref{e:z3}), (\ref{e:z5}), and (\ref{e:z6}), and
replacing the terms $\partial_t D_{cab}$ by $\partial_c\partial_t
g_{ab}/2$.  We note that these fields $Z^2_i$, $Z^3_i$, $Z^5_{ij}$ and
$Z^6_{kij}$ can also be controlled using the Type I boundary
conditions described in Sec.~\ref{s:TypeI}.  We have tested these Type
II boundary conditions numerically, and find the results to be
essentially identical to the results described in
Sec.~\ref{s:NumericalResults} for tests with Type I boundary
conditions.

%%%%%%%%%%%%%%%%%%%%%%%%%%%%%%%%%%%%%%%%%%%%%%%%%%%%%%%%%%%%%%%%%%%%%%%%%%%%%
\section{Physical Boundary Conditions}
\label{s:PhysicalBC}

In this section we derive boundary conditions for the physical
components of the dynamical fields, $U^{4-}_{ij}$, which represent the
real gravitational wave degrees of freedom of the system, at least
asymptotically.  Our strategy is based on the proposal of Bardeen and
Buchman~\cite{Bardeen2002}, and similar ideas used by Reula and
Sarbach~\cite{Reula2004} in the context of the Maxwell system.  These
ideas are adapted to the general 3D Einstein system by analyzing the
dynamics of the Weyl tensor as determined by the Bianchi
identities. We then infer boundary conditions on the dynamical fields of
the KST system that control the incoming radiative parts of the Weyl
tensor.  We begin by following the usual practice of decomposing the
Weyl tensor into electric and magnetic parts,
\begin{eqnarray}
E_{\mu\nu}&=&C_{\mu\sigma\nu\tau}\,T^\sigma T^\tau,\\
B_{\mu\nu}&=&\half
C_{\mu\omega\sigma\tau}\,\epsilon^{\sigma\tau}{}_{\nu\rho}T^\omega T^\rho,
\end{eqnarray}
where $T^\mu$ represents a timelike unit vector.  [In this paper
letters from the latter part of the Greek alphabet ($\mu$, $\nu$,
etc.) represent four-dimensional spacetime coordinate indices.]  In
our analysis here we assume that $T^\mu$ is orthogonal to the
$t=$constant spacelike hypersurfaces of the standard 3+1 decomposition
of the geometry.  In this case we can express the electric and
magnetic parts of the Weyl tensor in terms of standard 3+1 quantities:
\begin{eqnarray}
E_{ij}&=&R_{ij}+K K_{ij} -K_i{}^k K_{kj},\label{e:eijdef}\\
B_{ij}&=& \bigl(\nabla_k K_{li}\bigr)\epsilon^{kl}{}_j,\label{e:bijdef}
\end{eqnarray}
where $R_{ij}$, $K_{ij}$, $\nabla_i$ and $\epsilon_{ijk}$ represent
the three-dimensional Ricci tensor, the extrinsic curvature, the 3D
spatial covariant derivative, and the 3D totally antisymmetric tensor
respectively.  These electric and magnetic parts of the Weyl tensor
are symmetric and traceless in any vacuum spacetime.

The Weyl tensor satisfies the Bianchi identities,
\begin{eqnarray}
0=\nabla_{[\sigma}C_{\tau\omega]\mu\nu},
\end{eqnarray}
in any vacuum spacetime, where here $\nabla_\mu$ represents the 4D
covariant derivative, and these equations imply a system of evolution
equations for the Weyl tensor.  The 3+1 representation of the
principal parts of these evolution equations can be written~\cite{friedrich96}
in the form
\begin{eqnarray}
\partial_t E_{ij}-N^k\partial_k E_{ij}&\simeq & -N \bigl[
\partial_k B_{l(i}\bigr]\epsilon_{j)}{}^{kl},\\
\partial_t B_{ij}-N^k\partial_k B_{ij}&\simeq & N \bigl[
\partial_k E_{l(i}\bigr]\epsilon_{j)}{}^{kl},
\end{eqnarray}
which is reminiscent of the Maxwell system.  The
characteristic fields of this vacuum Weyl tensor evolution system
are
\begin{eqnarray}
Z^{13}&=& n^a n^b E_{ab},\\
Z^{14}&=& n^a n^b B_{ab},\\
U^{7\pm}_i&=& P^a{}_i n^b E_{ab} \mp \epsilon_{i}{}^{ab} n_b n^c
B_{ac},\\
U^{8\pm}_{ij}&=&\bigl(P^{(a}{}_i P^{b)}{}_j-\half P_{ij} P^{ab}\bigr)
\bigl(E_{ab}\mp \epsilon_{a}{}^{cd} n_d B_{cb}\bigr),\qquad\label{e:u8def}
\end{eqnarray}
where $n^k$ is a spatial unit vector.  The characteristic fields
$Z^{13}$ and $Z^{14}$ have characteristic speed zero, $U^{7\pm}_i$ have
speeds $\pm 1/2$, and $U^{8\pm}_{ij}$ have speeds $\pm 1$ relative
to the hypersurface orthogonal observers.  These fields propagate
with coordinate speeds $-n_kN^k$, $\pm N/2-n_kN^k$, and
$\pm N-n_kN^k$ respectively.  An equivalent expression for $U^{8\pm}_{ij}$
in terms of the standard 3+1 variables is
\begin{eqnarray}
U^{8\pm}_{ij}&=&\bigl(P^{a}{}_i P^{b}{}_j-\half P_{ij} P^{ab}\bigr)
\Bigl(R_{ab}+ K K_{ab} - K_a{}^c K_{cb}\nonumber\\
&&\qquad\qquad\qquad\mp n^c\nabla_c K_{ab} \pm n^c \nabla_{(a}K_{b)c}\Bigr).
\qquad\label{e:u8newdef}
\end{eqnarray}
The vacuum characteristic fields $U^{8+}_{ij}$ and $U^{8-}_{ij}$ are
proportional to the Newman-Penrose~\cite{Newman1962}
components of the Weyl tensor $\Psi_4$ and $\Psi_0$, respectively.

The true gravitational wave degrees of freedom are represented by the
fields $U^{8\pm}_{ij}$.  The best (gauge invariant) way to impose
boundary conditions on the physical degrees of freedom of the Einstein
system is therefore to require that the incoming part of this field
has prescribed values: $U^{8-}_{ij}=U^{8-}_{ij}|_{\hbox{bc}}$ at the
boundaries.
This condition is equivalent to fixing the Newman-Penrose
component $\Psi_0=\Psi_0|_{\hbox{bc}}$ on the boundary.  Since
$U^{8-}_{ij}$ falls to zero as $r^{-5}$ in an asymptotically flat
spacetime~\cite{Newman1962}, it is not unreasonable simply to set
$U^{8-}_{ij}=0$ as a physical boundary condition.  However, since
$U^{8-}_{ij}\neq 0$ in the Kerr geometry, we think it is more
consistent to freeze $U^{8-}_{ij}$ to its initial value by setting
$U^{8-}_{ij}|_{\hbox{bc}}=U^{8-}_{ij}(t\!=\!0)$.  

In order to write the boundary condition on $U^{8-}_{ij}$ as a condition
on the principal dynamical fields, we first
express $U^{8-}_{ij}$ in terms of the principal characteristic fields
using Eq.~(\ref{e:u8newdef}):
\begin{eqnarray}
&&\!\!\!\!\!\!\!
U^{8-}_{ij}= d_\perp U^{4-}_{ij} -\gamma_2 Z^{11}_{(ij)}
-\bigl[P^{(a}{}_i P^{b)}{}_j-\half P_{ij} P^{ab}\bigr]\nonumber\\
&&\times\biggl\{n^c\partial_a K_{bc} 
+ 2\gamma_2 P^{cd}\partial_{[a} D_{c]bd}+\gamma_2 n^cn^d\partial_a D_{cbd}
\nonumber\\
&&\quad+2P^{cd}\partial_c D_{[da]b}
+ 2 g^{cd}\partial_a D_{[bc]d} -K K_{ab}
+K_{ac}K^c{}_b\nonumber\\
&&\quad+(2D^{cd}{}_c-D^{dc}{}_c)(2D_{abd}-D_{dab}\bigr)
+4 D^{cd}{}_aD_{[dc]b}\,\,\,\nonumber\\
&&\quad- D_a{}^{cd}D_{bcd}
+2n^c K^d{}_{[a}D_{c]bd}+4n^c D_{[bd][c}K_{a]}{}^d\biggr\}.\nonumber\\
\end{eqnarray}
In deriving this expression, we have used the fact that the constraint
characteristic field $Z^{11}_{ij}$ can be expressed in terms of
$d_\perp Z^5_{ij}$ using Eq.~(\ref{e:z11cond}).  Setting
$U^{8-}_{ij}=U^{8-}_{ij}|_{\hbox{bc}}$ (and $Z^{11}_{(ij)}=0$) gives
us the desired Neumann-like boundary condition for the principal
characteristic field $U^{4-}_{ij}$:
\begin{eqnarray}
&&\!\!\!\!\!\!d_\perp U^{4-}_{ij} =U^{8-}_{ij}|_{\hbox{bc}}+
\bigl[P^{(a}{}_i P^{b)}{}_j-\half P_{ij} P^{ab}\bigr]
\biggl\{n^c\partial_a K_{bc} 
\nonumber\\
&&\quad+ 2\gamma_2 P^{cd}\partial_{[a} D_{c]bd}
+\gamma_2n^cn^d\partial_a D_{cdb}
+2P^{cd}\partial_c D_{[da]b} \nonumber\\
&&\quad+ 2 g^{cd}\partial_a D_{[bc]d}-K K_{ab}
+K_{ac}K^c{}_b+4 D^{cd}{}_aD_{[dc]b}\nonumber\\
&&\quad+(2D^{cd}{}_c-D^{dc}{}_c)(2D_{abd}-D_{dab}\bigr)
- D_a{}^{cd}D_{bcd}\,\,\,\nonumber\\
&&\quad+2n^c K^d{}_{[a}D_{c]bd}+4n^c D_{[bd][c}K_{a]}{}^d\biggr\}.
\end{eqnarray}
This physical boundary condition can be imposed numerically with the
same technique used for the Type I constraint-preserving boundary
conditions discussed in Sec.~\ref{s:TypeI}.  In particular we set
\begin{eqnarray}
&&\!\!\!\!\!\!d_t U^{4-}_{ij} = D_t U^{4-}_{ij}\nonumber\\
&&\qquad - (N + n_k N^k)\Bigl[U^{8-}_{ij} 
-U^{8-}_{ij}|_{\hbox{bc}}+ \gamma_2 Z^{11}_{(ij)}\Bigr].\qquad
\label{eq:physicalBc}
\end{eqnarray}
The term $D_t U^{4-}_{ij}$ is to be evaluated using Eq.~(\ref{e:Dtuhatdef}),
while $Z^{11}_{ij}$ and $U^{8-}_{ij}$ are to be evaluated numerically using
Eqs.~(\ref{e:z11def}) and (\ref{e:u8newdef}) respectively.

%%%%%%%%%%%%%%%%%%%%%%%%%%%%%%%%%%%%%%%%%%%%%%%%%%%%%%%%%%%%%%%%%%%%%%%%%%%%%%%
\section{Gauge Fixing Boundary Conditions}
\label{s:GaugeFixingBC}

Next we turn our attention to finding boundary conditions for the two
fields $U^{1-}$ and $Z^4_i$ that were not fixed by either the
constraint preserving boundary conditions in
Sec.~\ref{s:ConstraintPreservingBC} or the physical boundary
conditions in Sec.~\ref{s:PhysicalBC}.  Since these fields are not
fixed by the constraints or by physical considerations, they must be
gauge fields in effect.  Boundary
conditions on $U^{1-}$ and $Z^4_i$ should be chosen in a way that is
consistent with the gauge conditions.  In the
evolution system used here, we assume that the lapse density and
the shift are frozen to their initial values.  One reasonable
boundary condition for these gauge fields then is simply to freeze
them at their initial values on the boundaries:
\begin{eqnarray}
d_t U^{1-}&=& 0,\label{e:freezeu1}\\
d_t Z^4_i\;\; &=&0.\label{e:freezez4}
\end{eqnarray}

We have also found that a useful boundary condition for $U^{1-}$ can
be obtained from one of the standard equations used to fix the lapse:
$\partial_t K=0$.  Expressing $\partial_t K$ in terms of the time
derivatives of the principal characteristic fields, we find
\begin{eqnarray}
\partial_t K &=& g^{ij} \partial_t K_{ij} - K^{ij} \partial_t g_{ij},
\nonumber\\
&=& \frac{3-q}{4 v_3} \Bigl(d_t U^{3+} + d_t U^{3-}\Bigr)\nonumber\\
&&+ \frac{1}{4v_1}\Bigl(d_t U^{1+} + d_t U^{1-}\Bigr)
-K^{ij}\partial_t g_{ij}.\label{e:kdot}
\end{eqnarray}
[The parameter $q$ that appears in this expression is the ratio of
characteristic speeds defined in Eq.~(\ref{eq:CharFieldQdefinition})].
Setting $\partial_t K=0$ therefore provides a boundary condition for
the incoming characteristic field $U^{1-}$:
\begin{eqnarray}
d_t U^{1-} &= &- d_t U^{1+} - \frac{v_1(3-q)}{v_3} 
\bigl(d_t U^{3+} + d_t U^{3-}\bigr)\nonumber\\
&&+4 v_1 K^{ij}\partial_t g_{ij}.\label{e:kfreezebc}
\end{eqnarray}
The quantities $d_t U^{3-}$ and $\partial_t g_{ij}$ on the right side
of Eq.~(\ref{e:kfreezebc}) are evaluated with the appropriate boundary
expressions for these quantities.  We find that using
Eq.~(\ref{e:kfreezebc}) in evolutions of black-hole spacetimes is
fairly effective in controlling the growth of perturbations in
$U^{1-}$ with low spherical harmonic indices.  But at the same time
using this boundary condition causes unstable (and non-convergent)
growth of perturbations with large spherical harmonic indices.  We
solve this problem numerically by applying this boundary condition
only after applying a filter that removes all spherical harmonic
components above $\ell=2$ from Eq.~(\ref{e:kfreezebc}) at the
boundary.  Thus we use Eq.~(\ref{e:kfreezebc}) for the $\ell\leq 2$
spherical harmonic components of this equation, and a regular freezing
boundary condition, Eq.~(\ref{e:freezeu1}), for the $\ell>2$ spherical
harmonic components.

Finally, we note that boundary conditions for $U^{1-}$ and $Z^4_i$ can
also be obtained from the ``$\Gamma$-freezing'' condition that is
often used to determine the shift vector~\cite{Alcubierre2001,
Alcubierre2002}.  We found that these $\Gamma$-freezing boundary
conditions on $U^{1-}$ and $Z^4_i$ are not as effective as
Eqs.~(\ref{e:freezez4}) and (\ref{e:kfreezebc}) in controlling the
growth of instabilities in these ``gauge'' fields.  And since the
equations for the $\Gamma$-freezing forms of these boundary conditions
are quite lengthy, we do not reproduce them here.

%%%%%%%%%%%%%%%%%%%%%%%%%%%%%%%%%%%%%%%%%%%%%%%%%%%%%%%%%%%%%%%%%%%%%%%%%%%%%%%
\section{Numerical Methods}
\label{s:NumericalMethods}

In this section we describe briefly the numerical methods used to
compute the simulations presented below.  All our numerical
computations are performed using a multidomain pseudospectral
collocation method.  Our numerical methods are essentially the same as
those we applied previously to evolution problems with scalar
fields~\cite{Scheel2004, Holst2004}, with the Maxwell
system~\cite{Lindblom2004}, and the Einstein system~\cite{Kidder2000a,
Kidder2001, Lindblom2002, Scheel2002}.

%%%%%%%%%%%%%%%%%%%%%%%%%%%%%%%%%%%%%%%%%%%%%%%%%%%%%%%%%%%%%%%%%%%%%%%%%%%%%%%
\subsection{Spectral Collocation Method}
\label{s:NumericalMethodsSingleSubdomain}

The computational domain for the single black-hole evolutions
described in Sec.~\ref{s:NumericalResults} is a spherical shell
extending from some $r_{\mathrm{min}}$ (just inside the black-hole
event horizon) to some maximum value $r_\mathrm{max}$.  This domain
may also be subdivided into one or more spherical-shell subdomains.
Consider a single one of these subdomains that extends from radius
$r'_{\mathrm{min}}$ to $r'_{\mathrm{max}}$.  Given a system of partial
differential equations
\begin{equation} 
\partial_t u^\alpha(\mathbf{x},t) = 
{\cal F}^\alpha[u(\mathbf{x},t),\partial_i u(\mathbf{x},t) ],
\label{diffeq}
\end{equation}
where $u^\alpha$ is a collection of dynamical fields, the solution
$u^\alpha(\mathbf{x},t)$ on this subdomain is expressed as a
time-dependent linear combination of $N$ spatial basis functions
$\phi_k(\mathbf{x})$:
\begin{equation}
u^\alpha_N(\mathbf{x},t) = 
        \sum_{k=0}^{N-1}\tilde{u}^\alpha_k(t) \phi_k(\mathbf{x}).
\label{decom}
\end{equation}
We expand each Cartesian component of each tensor in terms of the
basis functions $T_n(\rho)Y_{\ell m}(\theta,\varphi)$, where $Y_{\ell
m}(\theta,\varphi)$ are spherical harmonics and $T_n(\rho)$ are
Chebyshev polynomials with
\begin{equation}
\rho=\frac{2r-r'_{\mathrm{max}}-r'_{\mathrm{min}}} 
         {r'_{\mathrm{max}}-r'_{\mathrm min}}.
\end{equation}
The spherical coordinates ($r$, $\theta$, and $\varphi$) used in these
spectral expansions are related to the pseudo-Cartesian coordinates
used to evaluate the components of tensor fields ({\it e.g.}
$g_{xx}$, $K_{yz}$, $D_{xyz}$, $...$) by the usual transformation
$x=r\sin\theta\sin\varphi$, $y=r\sin\theta\cos\varphi$, and
$z=r\cos\theta$.  In these spectral expansions we keep Chebyshev
polynomials $T_n(\rho)$ with $n$ up to some finite $N_r$, and
spherical harmonics $Y_{\ell m}(\theta,\varphi)$ with $\ell$ up to
some finite $\ell_{\mathrm{max}}$. For each $\ell$, we keep all
$|m|\le\ell$.  The values of $N_r$ and $\ell_{\mathrm{max}}$ determine
our radial and angular resolution, and we vary these in order to
perform convergence tests.  Spatial derivatives are evaluated
analytically using the known derivatives of the basis functions:
\begin{equation}
\partial_i u^\alpha_N(\mathbf{x},t) 
= \sum_{k=0}^{N-1}\tilde{u}^\alpha_k(t)
  \partial_i\phi_k(\mathbf{x}).
\label{decomderiv}
\end{equation}

Associated with the basis functions is a set of $N_c$ collocation
points $\mathbf{x}_i$.  Given spectral coefficients $\tilde
u^\alpha_k(t)$, the function values at the collocation points
$u^\alpha(\mathbf{x}_i,t)$ are computed using Eq.~(\ref{decom}).
Conversely, the spectral coefficients are obtained by the inverse
transform
\begin{equation}
\tilde{u}^\alpha_k(t) = \sum_{i=0}^{N_c-1} w_i u^\alpha_N(\mathbf{x}_i,t)
                       \phi_k(\mathbf{x}_i), 
\label{invdecom}
\end{equation}
where $w_i$ are weights specific to the choice of basis functions and
collocation points.  Thus it is straightforward to transform between
the spectral coefficients $\tilde{u}^\alpha_k(t)$ and the function
values at the collocation points $u^\alpha_N(\mathbf{x}_i,t)$.  The
partial differential equation, Eq.~(\ref{diffeq}), is now rewritten
using Eqs.~(\ref{decom})--(\ref{invdecom}) as a set of {\it
ordinary\/} differential equations for the function values at the
collocation points,
\begin{equation} 
\partial_t u^\alpha_N(\mathbf{x}_i,t) 
                      = {\cal G}^\alpha_i [u_N(\mathbf{x}_j,t)],
\label{odiffeq}
\end{equation}
where ${\cal G}^\alpha_i$ depends on $u^\alpha_N(\mathbf{x}_j,t)$ for
all $j$.  We integrate this system of ordinary differential equations,
Eq.~(\ref{odiffeq}), in time using a fourth-order Runge-Kutta
algorithm.  Boundary conditions are incorporated into the right side
of Eq.~(\ref{odiffeq}) using the technique of
Bj{\o}rhus~\cite{Bjorhus1995}. For the evolutions reported here the
time step is typically chosen to be about $1.5$ times the distance
between the closest collocation points in order that the
Courant-Friedrichs-Lewy stability limit remains satisfied.

We use no filtering on the radial basis functions, but apply a rather
complicated filtering rule for the angular functions.  When evaluating
the right side of Eq.~(\ref{odiffeq}), we set to zero the coefficients
of the terms with $\ell\geq\ell_{\mathrm{max}}-3$ in the tensor
spherical-harmonic expansions of the dynamical-field components ({\it
i.e.} the $g_{ij}$, $K_{ij}$, and $D_{kij}$ components) of this
equation.  This filtering method eliminates a certain type of angular
instability that arises because differentiation mixes the
various spherical-harmonic indices in the spectral expansions of
the Cartesian components of tensors.

%%%%%%%%%%%%%%%%%%%%%%%%%%%%%%%%%%%%%%%%%%%%%%%%%%%%%%%%%%%%%%%%%%%%%%%%%%%%%%%
\subsection{Multiple Subdomains}

Under many circumstances it is advantageous to divide the
computational domain into multiple subdomains.  
This subdivision allows faster calculations by
spreading the computational load over multiple processors.  It also
allows the use of different spectral resolutions in
different parts of the computational domain, thus making it possible
to concentrate computational resources in areas where they are most
needed.  We use the spectral collocation method described in
Sec.~\ref{s:NumericalMethodsSingleSubdomain} for each computational
subdomain.  An additional complication with multiple
subdomains is that only some of the subdomain boundaries are external, while
some are just internal boundaries that separate subdomains.
Appropriate boundary conditions must nevertheless be specified on each
boundary of each subdomain.  For the computations described here we
impose conditions on each subdomain boundary using the
Bj{\o}rhus~\cite{Bjorhus1995} method, in which $d_t u^{\hat\alpha}$
[see Eq.~(\ref{eq:dtudefinition})] is specified for each ingoing
characteristic field ({\it i.e.,} for each ${\hat\alpha}$ with
$v_{(\hat\alpha)}<0$).  Non-ingoing fields do not need and are not
given boundary conditions.  For external boundaries (those without
neighboring subdomains) the incoming $d_t u^{\hat\alpha}$ are
computed according to some externally imposed boundary condition, for
example our new constraint-preserving or physical boundary conditions.
For internal boundaries, the values of all incoming $d_t
u^{\hat\alpha}$ are simply copied from the corresponding outgoing $d_t
u^{\hat\alpha}$ in the neighboring subdomain.  At a black-hole
excision boundary $v_{(\hat\alpha)}>0$ for all ${\hat\alpha}$, so no
boundary condition is required there on any of the dynamical fields.

For the simulations described here, we subdivide our computational
domain into a set of concentric spherical-shell subdomains. We choose
the constants $N_r$ and $\ell_{\mathrm{max}}$ (that specify the
spectral resolution of each subdomain) to have the same values in all
the subdomains; this makes it easier to achieve load balancing when
multiple processors are used.  We concentrate the numerical
resolution where it is needed in our solutions by allowing 
spherical shells with varying thickness: for fixed radial
resolution $N_r$, thin shells achieve higher resolution than thick
shells.

%%%%%%%%%%%%%%%%%%%%%%%%%%%%%%%%%%%%%%%%%%%%%%%%%%%%%%%%%%%%%%%%%%%%%%%%%%%%%%%
\subsection{Residual Evaluators}
\label{s:ResidualEvaluators}

We use three different computational diagnostics to help us evaluate the
accuracy and stability of the numerical simulations described in
Sec.~\ref{s:NumericalResults}.  The first and simplest diagnostic just
measures the difference between a numerical solution and the exact
solution of the equations.  This is possible whenever the exact
solution to the problem is known, as in our evolutions of unperturbed
black-hole spacetimes.  We measure the deviation of the numerical
solution from the exact solution quantitatively by evaluating the energy norm
of the difference between the two solutions:
\begin{eqnarray}
(\delta\!E)^2 = \int S_{\alpha\beta}(u^\alpha-u^\alpha_0)(u^\beta-u^\beta_0)
\sqrt{g}\, d^{\,3}x.\quad\label{e:energynorm}
\end{eqnarray}
Here $u^\alpha$ denotes the numerical solution, $u^\alpha_0$ the exact
solution, and $S_{\alpha\beta}$ the symmetrizer matrix of the
hyperbolic evolution system. The symmetrizer $S_{\alpha\beta}$ is a
positive definite matrix on the space of dynamical fields, which is
derived for the KST evolution system studied here in
Refs.~\cite{Lindblom2002, Lindblom2003}.  (We evaluate
$S_{\alpha\beta}$ using the exact solution $u^\alpha_0$.)  The energy
$\delta\!E$ is not dimensionless, and consequently it is not clear how to
interpret whether a given value means that the solution is a good
approximation of the exact solution or not.  Therefore we normalize by
dividing by the total energy of the exact solution:
\begin{eqnarray}
E^2_0 = \int S_{\alpha\beta}u^\alpha_0u^\beta_0
\sqrt{g}\, d^{\,3}x.\label{e:energydenom}
\end{eqnarray}
The ratio $\delta\!E/E_0$ is therefore a good dimensionless measure of the
accuracy of our numerical solution.  When this ratio becomes of order
unity, the numerical solution bears little resemblance to the exact
solution.

A second, more general, method of measuring the accuracy and
stability of our numerical solutions is to monitor the magnitudes of
the constraints, $c^A=\{{\cal C},{\cal C}_i,{\cal C}_{kij},{\cal
C}_{klij}\}$.  When the constraints vanish, the solutions to our first
order evolution system, Eqs.~(\ref{e:kstggauge})--(\ref{e:kstdgauge}),
are guaranteed to be solutions of the original Einstein equations as
well.  Thus, to measure how well our numerical solutions solve the
original Einstein system we construct the following measure of the
constraints:
\begin{eqnarray}
||\kernedcalC||^2 = \int\!\! \left({\cal C}^2 +{\cal C}_i{\cal C}^i +
{\cal C}_{kij}{\cal C}^{kij}+ {\cal C}_{klij}{\cal C}^{klij}\right)
\!\sqrt{g}\, d^{\,3}x.\,\,\,\label{e:constraintnorm}
\end{eqnarray}
The constraints consist of various derivatives of the dynamical fields
$u^\alpha$.  Therefore we can construct a meaningful dimensionless
measure of the constraints by normalizing $||\kernedcalC||$ by
dividing by a measure of these derivatives:
\begin{eqnarray}
||\partial u||^2 
= \int \sum_\alpha g^{ij}\,\partial_i u^\alpha
\partial_ju^\alpha \sqrt{g}\, d^{\,3}x.\label{e:dunorm}
\end{eqnarray}
We evaluate the quantities that appear in Eq.~(\ref{e:dunorm}) ({\it
i.e.,} the tensor components of $u^\alpha$ and the partial
derivatives) in the global Cartesian-like coordinate system used in
our code.  The ratio $||\kernedcalC||/||\partial
u||$ is therefore a dimensionless measure of the degree to which our
numerical solutions satisfy the original Einstein system.  When this
quantity becomes of order unity the constraint violations dominate
and our numerical solutions no longer accurately represent solutions
to the original Einstein equations.

At the analytical level, the vanishing of $||\kernedcalC||$ is
enough to ensure that the solutions to our first-order evolution
system also represent solutions to the original second-order Einstein
equations.  At the numerical level, however, there remains a (small we
think) possibility that our solutions to the discrete first-order
evolution system manage to keep $||\kernedcalC||$ small while
somehow failing to satisfy the original second-order Einstein system.
In order to detect whether this has occurred, we construct a third
independent measure of the degree to which our numerical solutions
satisfy the original system.  Following the ideas of
Choptuik~\cite{Choptuik1994}, we use our numerical solution for
$u^{\alpha}$ to reconstruct the four-dimensional metric:
\begin{eqnarray}
ds^2&=&g_{\mu\nu}dx^\mu dx^\nu\nonumber\\
&=&-N^2 dt^2 + g_{ij}(dx^i + N^i dt)(dx^j + N^j dt).\qquad
\label{e:fourmetricdef}
\end{eqnarray}
[Letters from the latter part of the Greek alphabet ($\mu$, $\nu$,
etc.) denote four-dimensional spacetime indices in our Cartesian-like
coordinate system.]  We evaluate the components of $g_{\mu\nu}$ on the
actual time slices used in our evolution code, and then evaluate the
time derivatives $\partial_t g_{\mu\nu}$ and $\partial_t^2 g_{\mu\nu}$
using standard centered finite-difference expressions for these
quantities with seven-point stencils.  Given these time derivatives on
a certain time slice, we then evaluate the complete set of first and
second derivatives, $\partial_\sigma g_{\mu\nu}$ and
$\partial_\sigma\partial_\tau g_{\mu\nu}$, by computing the spatial
derivatives using the spectral methods described in
Sec.~\ref{s:NumericalMethodsSingleSubdomain}.  Finally, we combine
these first and second derivatives to determine the four-dimensional
Ricci tensor:
\begin{eqnarray}
R_{\mu\nu}&=&-\half g^{\sigma\tau}\left(
\partial_\sigma\partial_\tau g_{\mu\nu}
+\partial_\mu\partial_\nu g_{\sigma\tau}
-2\partial_\sigma\partial_{(\mu}g_{\nu)\tau}\right)\nonumber\\
&&+\fourth g^{\sigma\tau}g^{\omega\rho}\Bigl[
\partial_\mu g_{\sigma\omega}\partial_\nu g_{\tau\rho}
+4\partial_\sigma g_{\mu\omega}\partial_{[\tau}g_{\rho]\nu}\nonumber\\
&&-\bigl(\partial_\omega g_{\mu\nu}
-2\partial_{(\mu}g_{\nu)\omega}\bigr)
\bigl(\partial_\rho g_{\sigma\tau} - 2 \partial_\sigma g_{\tau\rho}\bigr) 
\Bigr].\label{e:Ricci}
\end{eqnarray}
The Ricci tensor should vanish for any solution of the vacuum Einstein
equations, so this quantity gives us an independent way to measure how
well our numerical solution solves the original second-order Einstein
equations.

The expression for the Ricci tensor, Eq.~(\ref{e:Ricci}), has thirteen
terms (each of which contains contractions): four terms proportional
to $\partial_\sigma\partial_\tau g_{\mu\nu}$, and nine terms
proportional to $\partial_\sigma g_{\mu\nu}$ (when symmetrizations and
antisymmetrizations are expanded out).  We construct a positive
definite quantity, $R^{\scriptscriptstyle\mathrm{
RMS}}_{\mu\nu}$, to which the size of $R_{\mu\nu}$ can be
compared, by taking the square root of the sum of the squares of these
thirteen terms (for each $\mu$ and $\nu$).  We then construct
norms of these quantities by integrating them over our spacelike
slices:
\begin{eqnarray}
||{R}\kern 0.1em||^2&=& 
\int \sum_{\mu\leq\nu} (R_{\mu\nu})^2 \sqrt{g}\, 
d^{\,3}x,\label{e:Ricciresidual}\\
||\kern0.1em{R^{\scriptscriptstyle\mathrm{ RMS}}} ||^2 
&=& \int \sum_{\mu\leq\nu} 
(R^{\scriptscriptstyle\mathrm{ RMS}}_{\mu\nu})^2 \sqrt{g}\, d^{\,3}x.
\end{eqnarray}
We note that the $R^{\scriptscriptstyle\mathrm{ RMS}}_{\mu\nu}$
and $R_{\mu\nu}$ used to compute these norms are the
quasi-Cartesian components of these quantities used in our code.  We
use the ratio $||R||/||R^{\scriptscriptstyle RMS}||$ as our third
independent measure of the degree to which our numerical solutions
satisfy the original Einstein equations.  When both
$||R||/||R^{\scriptscriptstyle\mathrm{ RMS}}||$ and 
$||\kernedcalC||/||\partial u||$ are small, we have
considerable confidence in the accuracy of our solutions even in cases
where the exact solution is not known.

%%%%%%%%%%%%%%%%%%%%%%%%%%%%%%%%%%%%%%%%%%%%%%%%%%%%%%%%%%%%%%%%%%%%%%%%%%%%%%%
\section{Numerical Evolutions}
\label{s:NumericalResults}

We have studied the efficacy of the boundary conditions proposed in
Secs.~\ref{s:ConstraintPreservingBC}--\ref{s:GaugeFixingBC} by
performing evolutions of various single black-hole spacetimes.  These
simulations evolve either unperturbed Schwarzschild black holes in
Kerr-Schild coordinates,
\begin{eqnarray}
ds^2 = -dt^2 + \frac{2M}{r}(dt+dr)^2 + dr^2 + r^2 d\Omega^2,
\label{e:KerrSchildMetric}
\end{eqnarray}
or fully dynamical black holes that are small perturbations of the
Schwarzschild spacetime (but we solve the full nonlinear
equations in our code).  We express all lengths and times associated
with these simulations in units of the bare black-hole mass $M$, even
for the perturbed solutions in which the ADM mass exceeds $M$.  The
computational domains for these evolutions consist of one or more
concentric spherical shells that cover the space from
$r_{\mathrm{min}}=1.9M$ (just inside the black-hole event horizon) to
some maximum value $r_\mathrm{max}$.  We have run evolutions using several
different $r_\mathrm{max}$ in the range $6.9M\leq r_\mathrm{max}\leq
41.9M$.

\begin{figure} 
\begin{center}
\includegraphics[width=3in]{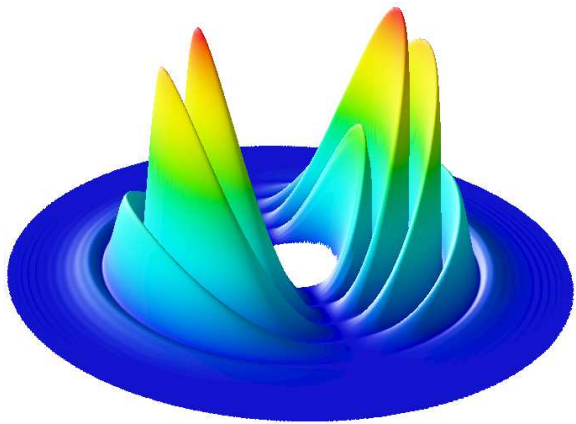}
\end{center}
\caption{Initial data for the perturbed black-hole evolutions.  Shown
is the Newman-Penrose component of the Weyl tensor $|\Psi_4|$,
which measures the outgoing gravitational wave flux.
\label{Psi4InitialData}}
\end{figure}

The initial data for these evolutions are prepared by applying an
odd-parity outgoing quadrupole-wave perturbation with amplitude
$4\times 10^{-3}$ to the Kerr-Schild spatial metric and its time
derivative.  This outgoing wave pulse is constructed by Teukolsky's
method~\cite{teukolsky82} with the generating function $G(r)$ that
determines the radial profile of the wave: $G(r)=Ae^{-(r-r_0)^2/w^2}$,
where $A=4\times10^{-3}$, $r_0=5M$ and $w=1.5M$.  Then we solve
numerically the full non-linear conformal ``thin-sandwich'' form of
the initial value equations to obtain constraint-satisfying perturbed
black-hole initial data (see Pfeiffer, {\it et
al.\/}~\cite{Pfeiffer2004}). In the resulting initial data set, the
ADM energy exceeds the apparent horizon mass by about $10^{-5}M$.
Figure~\ref{Psi4InitialData} illustrates the initial data for these
perturbed black-hole evolutions.  The quantity shown in
Fig.~\ref{Psi4InitialData} is a measure of the outgoing gravitational
wave flux, $\sqrt{U^{8+}_{ij}U^{8+}_{kl}g^{ik}g^{jl}/2}$, defined
in Eq.~(\ref{e:u8def}). This quantity is equivalent to the
Newman-Penrose component of the Weyl tensor $|\Psi_4|$.  The
outer radius of the computational domain for the initial data shown in
Fig.~\ref{Psi4InitialData} is $r_\mathrm{max}=11.9M$.
%
%--------------------BEGIN--FIGURE-----------------------------------------
%---------------------END--FIGURE-----------------------------------------

For simplicity we use fixed gauge evolutions: the lapse density
$Q$ and the shift $N^i$ are set to their initial Kerr-Schild
values, Eq.~(\ref{e:KerrSchildMetric}), for all times.  For the
perturbed initial data, we also set $Q$ and $N^i$ to the
unperturbed analytical Kerr-Schild values, ignoring the solutions for
these fields given by the conformal ``thin-sandwich'' initial value equations.
Although more sophisticated gauge conditions should result in better
long-term evolutions, these simple fixed-gauge conditions are
adequate for the tests of the new constraint preserving boundary
conditions, which are of primary interest here.

The KST evolution system has a number of freely specifiable
parameters.  We set the KST parameters to the values $\gamma_0=0.5$,
$\gamma_1=-12$, $\gamma_2=-1$, $\gamma_3=0.16$, and $\gamma_4=-0.96$
for all the evolutions discussed here.  This choice of parameters is
one of the special cases that leave the parameter $q$ of
Eq.~(\ref{eq:CharFieldQdefinition}) ill-defined.  In this case the
choice of $q$ is arbitrary, and we set $q=1$ in all of the evolutions
discussed here (since this yields better performance than $q=0$).
For these parameter choices, both the fundamental
evolution equations and the constraint evolution system are symmetric
hyperbolic, and all characteristic speeds (relative to hypersurface
orthogonal observers) are either $0$ or $\pm 1$; in particular,
$v^2_1=v^2_2=v^2_3=1$.  The values of the KST parameters used here are
the same as those used in previous studies~\cite{Lindblom2002,
Scheel2002}. However, the current evolution equations are rather different
from those used in Refs.~\cite{Lindblom2002, Scheel2002}, in which
the evolution equations were modified by a kinematical change of
variables and by the addition of terms proportional to the constraint
${\cal C}_{kij}$.  Here we perform no such modification; that is, the
kinematical parameters defined in Refs.~\cite{Kidder2001,
Lindblom2002, Scheel2002} are chosen here to have the values
$\hat{k}=1$, $\hat{z} =\hat{a} =\hat{b} =\hat{c} =\hat{d} =\hat{e}=0$.
The values of the KST parameters in Refs.~\cite{Lindblom2002,
Scheel2002} were chosen to minimize the growth rate of the constraints
for evolutions of a single black hole in Painlev\'e-Gullstrand
coordinates.  Because we use a rather different set of evolution
equations, and because we evolve black holes in Kerr-Schild
coordinates, we do not expect that these parameter values
are optimal for our current evolutions.

%%%%%%%%%%%%%%%%%%%%%%%%%%%%%%%%%%%%%%%%%%%%%%%%%%%%%%%%%%%%%%%%%%%%%%%%%%%%%
\subsection{Unperturbed Black Holes}
\label{s:Schwarzschild}

In this subsection we describe three numerical tests involving 3D
evolutions of unperturbed Schwarzschild black holes.  These evolutions
use the Schwarzschild spatial metric and extrinsic curvature in
Kerr-Schild coordinates, Eq.~(\ref{e:KerrSchildMetric}), as initial
data.  The first and second tests explore the evolutions of these
black holes using simple ``freezing'' boundary conditions.  Freezing
boundary conditions set the time derivatives of each of the incoming
characteristic fields to zero at the boundaries, {\it i.e.},
\begin{equation}
d_t u^{\hat\alpha} = 0,\label{e:freezingbcdef}
\end{equation}
for all ${\hat\alpha}$ with $v_{(\hat\alpha)}<0$ [where $d_t
u^{\hat\alpha}$ is defined in Eq.~(\ref{eq:dtudefinition})].  These
boundary conditions are known to make the evolution equations
mathematically well-posed, but they are also known to do a poor job of
preventing the influx of constraint violations into the computational
domain for fully dynamical time-dependent
solutions~\cite{Lindblom2004, Holst2004}.  Freezing boundary
conditions, however, have been used rather successfully for time
independent solutions such as the unperturbed black-hole spacetimes
considered in this subsection~\cite{Lindblom2002, Scheel2002}.

\begin{figure} 
\begin{center}
\includegraphics[width=3in]{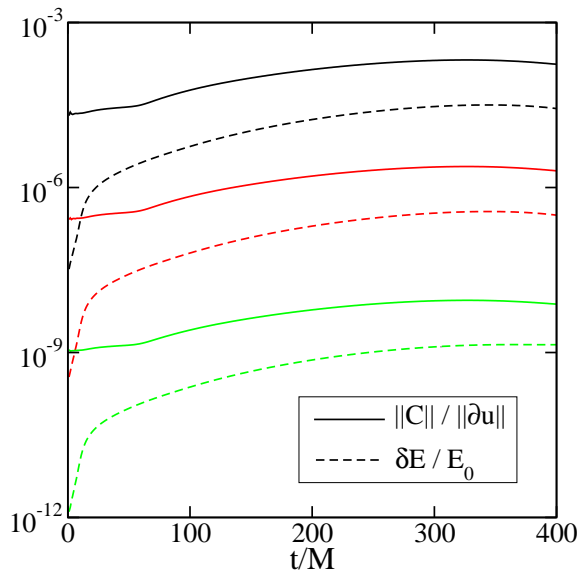}
\end{center}
\caption{Constraint violations and energy norms for unperturbed
black-hole evolutions using freezing boundary conditions. The outer
boundary is at $r_\mathrm{max}=41.9M$; and three different radial
resolutions, $N_r=17$, $21$, and $26$, show numerical convergence.
\label{Oct21CError}}
\end{figure}

Our first test evolves an unperturbed Schwarzschild black hole on a
computational domain extending from $r_\mathrm{min}=1.9M$ to
$r_\mathrm{max}=41.9M$.  This domain is subdivided into eight
subdomains, each of width $5M$, each using the same angular resolution
$\ell_\mathrm{max}=21$, and each having the same radial resolution
$N_r=17$, $21$ or $26$.  This angular resolution is much higher than
necessary to resolve this spherically symmetric spacetime, but we
wanted to verify the stability of our code even for large values of
$\ell_\mathrm{max}$.  Figure~\ref{Oct21CError} shows two measures of
the errors in these evolutions: $||\kernedcalC||/||\partial u||$ and 
$\delta\!E/E_0$.
These results show that our error measures converge toward zero as we
increase the radial resolution $N_r$.  These results also show
that our computational methods are capable of evolving unperturbed
black-hole solutions for hundreds (if not thousands) of $M$, which is
consistent with previous results using a similar evolution
system~\cite{Lindblom2002, Scheel2002}.
%
%--------------------BEGIN--FIGURE-----------------------------------------
%---------------------END--FIGURE-----------------------------------------

\begin{figure} 
\begin{center}
\includegraphics[width=3in]{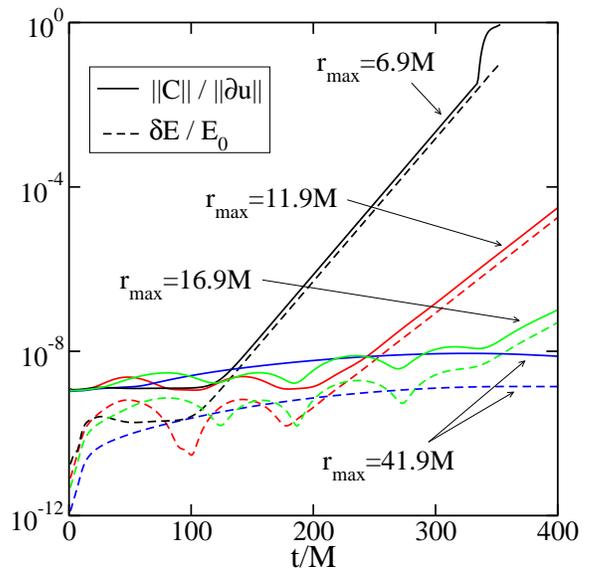}
\end{center}
\caption{Constraint violations and energy norms for unperturbed
black-hole evolutions with a range outer boundary radii,
$r_\mathrm{max}=6.9M$, $11.9M$, $16.9M$, $41.9M$.  These runs all use
$N_r=26$ in each subdomain, but use different numbers of subdomains to
achieve different $r_\mathrm{max}$.  They all use freezing boundary
conditions, and $\ell_\mathrm{max}=21$.
\label{UnpertFreezingOB}}
\end{figure}

\begin{figure} 
\begin{center}
\includegraphics[width=3in]{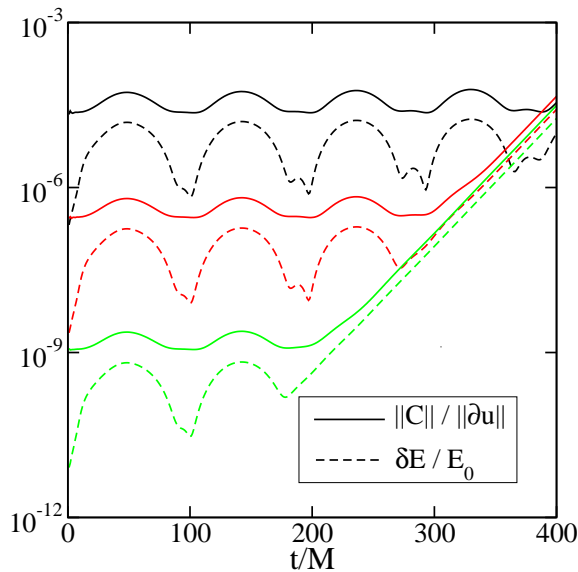}
\end{center}
\caption{Constraint violations and energy norms for an unperturbed
black-hole evolution using freezing boundary conditions.  The outer
boundary is at $r_\mathrm{max}=11.9M$; and three different radial
resolutions, $N_r=17$, $21$, and $26$, show numerical convergence
to an exponentially growing solution. 
\label{Oct21AError}}
\end{figure}

As a second test, we explore the effects of changing the location of
the outer boundary of the computational domain, $r_\mathrm{max}$, for
these unperturbed black-hole evolutions with freezing boundary
conditions.  Figure~\ref{UnpertFreezingOB} shows the constraint
violations $||\kernedcalC||/||\partial u||$ and the energy norms 
$\delta\!E/E_0$ for
evolutions using several different outer radii: $r_\mathrm{max}=6.9M$,
$11.9M$, $16.9M$, and $41.9M$.  All of these evolutions use the same
$\ell_\mathrm{max}=21$, and $N_r=26$ in each subdomain.  Each
computational subdomain has width $5M$, so the number of subdomains is
adjusted to achieve the desired $r_\mathrm{max}$.
Figure~\ref{UnpertFreezingOB} shows that these evolutions have an
instability that causes $||\kernedcalC||/||\partial u||$ and $\delta\!E/E_0$
to grow exponentially. This instability becomes weaker as $r_\mathrm{max}$
increases; we find (by measuring the growth rate for
$r_\mathrm{max}=6.9M$, $11.9M$, $16.9M$, $21.9M$, and $31.9M$)
that the growth rate is $M/\tau \approx e^{-r_\mathrm{max}/13M}$.
So this appears to be a
constraint-violating instability that is influenced by the location of
the outer boundary of the computational domain.
Figure~\ref{Oct21AError} demonstrates for the $r_\mathrm{max}=11.9M$
case that these unstable evolutions are numerically convergent, by
showing that $||\kernedcalC||/||\partial u||$ and $\delta\!E/E_0$ approach an
exponentially growing solution as $N_r$ increases.
Figure~\ref{Oct21AError} also shows that the growth rate of the
instability is the same for all values of $N_r$ in this
$r_\mathrm{max}=11.9M$ case.  (Numerical convergence and independence
of the growth rate with $N_r$ is also observed for the other values of
$r_\mathrm{max}$ that we tested.)  We have also verified that the
growth rate of this instability does not depend on the angular
resolution $\ell_\mathrm{max}$ for $7\le\ell_\mathrm{max}\le 21$.
These checks suggest that this constraint violating instability is a
solution of the continuum differential equations, and is not primarily
an artifact of the discrete numerical representation of the equations
used here.
%
%--------------------BEGIN--FIGURE-----------------------------------------
%---------------------END--FIGURE-----------------------------------------
%
%--------------------BEGIN--FIGURE-----------------------------------------
%---------------------END--FIGURE-----------------------------------------

\begin{figure} 
\begin{center}
\includegraphics[width=3in]{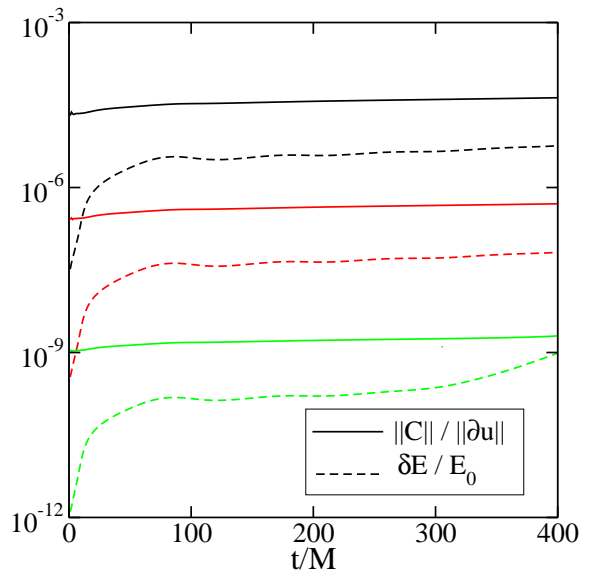}
\end{center}
\caption{Constraint violations and energy norms for unperturbed
black-hole evolutions using our new boundary conditions. The outer
boundary is at $r_\mathrm{max}=41.9M$; and three different radial
resolutions, $N_r=17$, $21$, and $26$, show numerical convergence.
\label{Oct21FError}}
\end{figure}

For the third numerical test we use our new boundary conditions to
evolve the same unperturbed black-hole initial data on the same
computational domains used in the first two tests.  These new boundary
conditions include the new constraint preserving boundary conditions,
Eqs.~(\ref{e:gdotbc})--(\ref{e:u3mbc}) with $\mu_5=0.75$ and
$\mu_6=-0.5$, the new physical boundary condition,
Eq.~(\ref{eq:physicalBc}), and the new gauge boundary conditions,
Eqs.~(\ref{e:freezez4}) and (\ref{e:kfreezebc}).
Figure~\ref{Oct21FError} depicts $||\kernedcalC||/||\partial u||$ and 
$\delta\!E/E_0$
for evolutions of the unperturbed black-hole initial data on a
computational domain with $r_\mathrm{max}=41.9M$.  (Except for
boundary conditions, the evolutions in Fig.~\ref{Oct21FError} are
identical to those of Fig.~\ref{Oct21CError}.)  The evolutions of
Fig.~\ref{Oct21FError} show that both the constraints and the energy
norms of these solutions decrease toward zero as $N_r$ increases.
%
%--------------------BEGIN--FIGURE-----------------------------------------
%---------------------END--FIGURE-----------------------------------------

\begin{figure} 
\begin{center}
\includegraphics[width=3in]{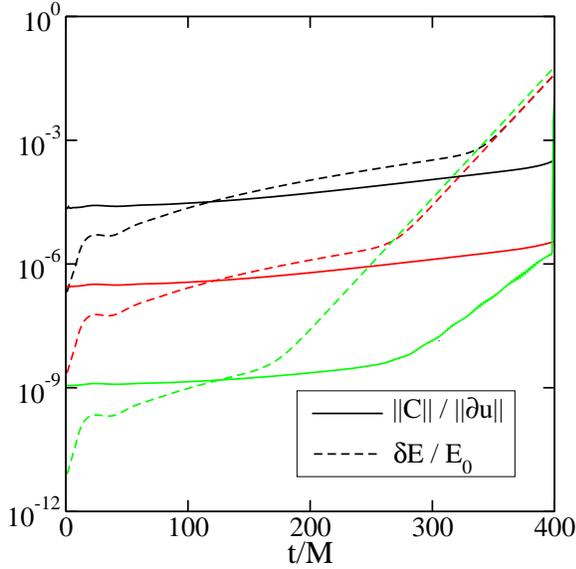}
\end{center}
\caption{Constraint violations and energy norms for an unperturbed
black-hole evolution using our new boundary conditions.  The outer
boundary is at $r_\mathrm{max}=11.9M$; and three different radial
resolutions, $N_r=17$, $21$, and $26$, show numerical convergence
to an exponentially growing solution.
\label{Oct21DError}}
\end{figure}

At late times, the highest resolution curve plotted in
Fig.~\ref{Oct21FError} shows the beginning of an exponentially-growing
mode.  We call this a gauge mode, because $\delta\!E/E_0$ begins to grow in
these solutions without a corresponding growth in the constraints
$||\kernedcalC||/||\partial u||$.  
This gauge mode grows more rapidly if we move
the outer boundary inward, as can be
seen in Fig.~\ref{Oct21DError}, which shows the analogous
evolutions with outer boundary at $r_\mathrm{max}=11.9M$.  The
constraints remain roughly constant in the evolutions of
Fig.~\ref{Oct21DError} for a considerable amount of time after $\delta\!E/E_0$
begins to grow rapidly.  Comparing the results of
Fig.~\ref{Oct21AError} with those of Fig.~\ref{Oct21DError}, we see
that our constraint-preserving boundary conditions do improve the
constraint violations in these evolutions, but only slightly.  However,
this improvement comes at the expense of introducing a new gauge-mode
instability.  At very late times we see that the highest resolution
evolution also shows signs of a constraint violating instability,
although it grows more slowly than the gauge mode.  The growth rate of
the constraints seen in the highest resolution evolution of
Fig.~\ref{Oct21DError} is the same as the growth rate of the
instability seen in Fig.~\ref{Oct21AError}, to within about $5\%$.
This suggests that both of these constraint violating instabilities
might be caused primarily by bulk rather than
boundary constraint violations.
%
%--------------------BEGIN--FIGURE-----------------------------------------
%---------------------END--FIGURE-----------------------------------------

The gauge-mode instability seen in Figs.~\ref{Oct21FError}
and~\ref{Oct21DError} is dominated by its $\ell=3$ spherical harmonic
component.  Recall that we impose the gauge-fixing boundary condition,
Eq.~(\ref{e:kfreezebc}), by filtering out everything but the $\ell\leq 2$
components of $d_tU^{1-}$.  If we change this filtering to impose
Eq.~(\ref{e:kfreezebc}) on only the $\ell\leq 1$ components of $d_tU^{1-}$,
the gauge-mode instability is dominated by its $\ell=2$ spherical
harmonic component, and grows more rapidly than in
Figs.~\ref{Oct21FError} and~\ref{Oct21DError}.  Thus the gauge
boundary condition, Eq.~(\ref{e:kfreezebc}), does improve the
stability of the gauge mode at least to some degree.  However, if we
impose Eq.~(\ref{e:kfreezebc}) without any filtering, the instability
grows much faster than the rate seen in Figs.~\ref{Oct21FError}
and~\ref{Oct21DError}; and this growth rate increases as
$\ell_\mathrm{max}$ increases, so the evolution becomes
non-convergent.  Recall that we use the simplest possible gauge
conditions: we set the lapse density $Q$ and the shift $N^i$
to their analytic values throughout the evolution.  A more
sophisticated treatment of gauge conditions (not just at the
boundaries, but throughout the volume) is probably needed to control
these unstable gauge modes.

%%%%%%%%%%%%%%%%%%%%%%%%%%%%%%%%%%%%%%%%%%%%%%%%%%%%%%%%%%%%%%%%%%%%%%%%%%%%%%%
\subsection{Perturbed Black Holes}
\label{sec:pert-schw-black}

In this subsection we describe 3D numerical evolutions of perturbed
black-hole spacetimes.  The initial data for these evolutions
are discussed at the beginning of Sec.~\ref{s:NumericalResults}.
We evolve these initial data using the same non-linear equations and
evolution methods used in Sec.~\ref{s:Schwarzschild}
for the unperturbed cases.  These perturbed black-hole initial data
include a short-wavelength gravitational-wave packet, so more radial
collocation points are required to achieve an accuracy comparable to
that of the unperturbed black-hole cases.

\begin{figure} 
\begin{center}
\includegraphics[width=3in]{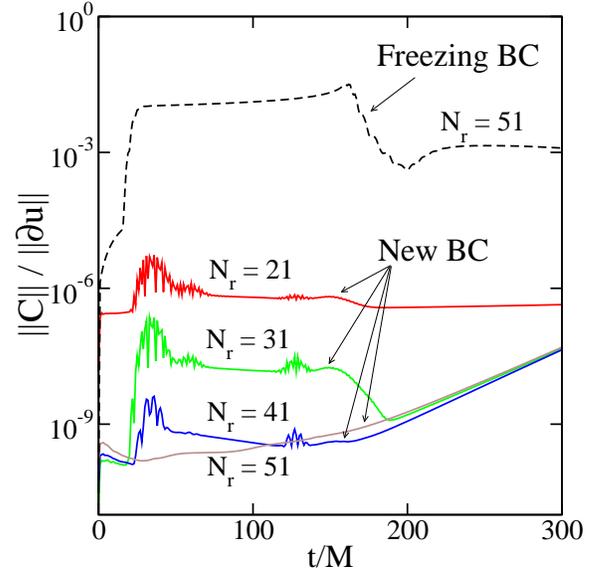}
\end{center}
\caption{Constraint violations for evolutions of perturbed black holes
using freezing (dashed curve) and our new boundary conditions (solid
curves).  Radial resolutions $N_r=21$, $31$, $41$, and $51$ are shown
for the new boundary conditions, while only $N_r=51$ is shown for the
freezing case.  The outer boundary is at $r_\mathrm{max}=21.9$.
\label{Nov3BvsNov9A}}
\end{figure}

Figure~\ref{Nov3BvsNov9A} shows the constraint error
$||\kernedcalC||/||\partial
u||$ for evolutions of these perturbed black-hole initial data.  These
simulations are performed on a computational domain with four
concentric subdomains, each of width $5M$ and $\ell_\mathrm{max}=11$,
and with outer boundary at $r_\mathrm{max}=21.9$.  The dashed curve in
Fig.~\ref{Nov3BvsNov9A} shows the results of using the simple freezing
boundary conditions, Eq.~(\ref{e:freezingbcdef}): a large constraint
violation is generated at $t\approx 20M$ when the wave pulse passes
through the outer boundary of the computational domain.  The size of
this constraint violation is comparable to the amplitude of the wave,
and does not converge away with increased resolution.  Only the
highest radial resolution is shown, because the lower resolution
curves are almost identical.  Therefore, these simulations with
freezing boundary conditions fail and can not be used to model the
physical gravitational waves in these solutions after the time
$t\approx 20M$ with any accuracy.
%
%--------------------BEGIN--FIGURE-----------------------------------------
%---------------------END--FIGURE-----------------------------------------

\begin{figure} 
\begin{center}
\includegraphics[width=3in]{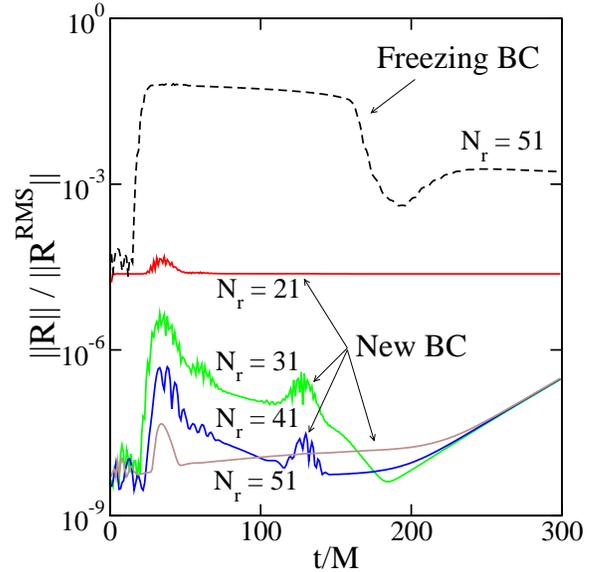}
\end{center}
\caption{Four-dimensional Ricci-tensor residual of
Eq.~(\ref{e:Ricciresidual}) for the perturbed black-hole evolutions
shown in Fig.~\ref{Nov3BvsNov9A}.
\label{Nov3BvsNov9ARicci}}
\end{figure}

Also plotted in Fig.~\ref{Nov3BvsNov9A} are evolutions of the same
perturbed black-hole initial data on the same computational domain,
but now using our new boundary conditions: the new constraint
preserving boundary conditions, Eqs.~(\ref{e:gdotbc})--(\ref{e:u3mbc})
with $\mu_5=0.75$ and $\mu_6=-0.5$, the new physical boundary
condition, Eq.~(\ref{eq:physicalBc}), and the new gauge boundary
conditions, Eqs.~(\ref{e:freezez4}) and (\ref{e:kfreezebc}).  In these
cases the constraint violation generated as the wave passes through
the outer boundary converges away with increasing radial resolution.
The constraint violation is eight orders of magnitude smaller than
with freezing boundary conditions in the highest resolution case.
Figure~\ref{Nov3BvsNov9ARicci} shows an independent measure of how
well these numerical evolutions satisfy the original Einstein
equations by plotting the average value of the four-dimensional Ricci
tensor $||R||/||R^{\mathrm{RMS}}||$ [see Eq.~(\ref{e:Ricciresidual})]
for the same simulations shown in Fig.~\ref{Nov3BvsNov9A}.
Figure~\ref{Nov3BvsNov9ARicci} confirms the results seen in
Fig.~\ref{Nov3BvsNov9A}: the Einstein equations are violated when the
wave hits the boundary with freezing boundary conditions, while the
new boundary conditions are quite effective in reducing this violation
by many orders of magnitude.
%
%--------------------BEGIN--FIGURE-----------------------------------------
%---------------------END--FIGURE-----------------------------------------

\begin{figure} 
\begin{center}
\includegraphics[width=3in]{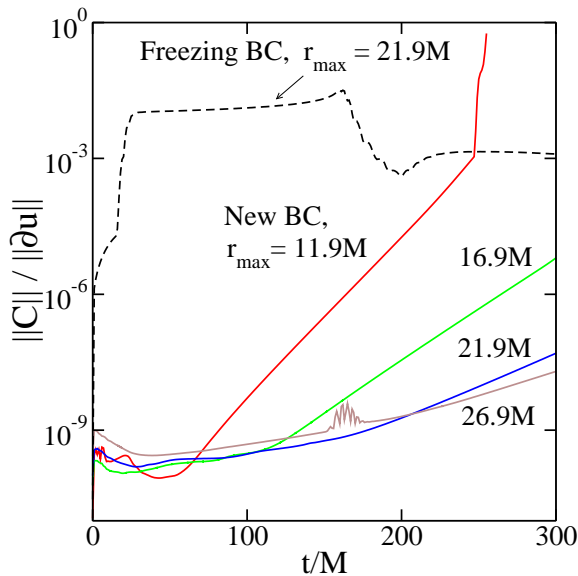}
\end{center}
\caption{Constraint violations in evolutions of a perturbed black
hole. The dashed curve corresponds to freezing boundary conditions, and
the solid curves correspond to the new boundary conditions for
different $r_\mathrm{max}$.
\label{ConstraintVsOuterBc}}
\end{figure}

Two interesting features are seen in the evolutions in
Figs.~\ref{Nov3BvsNov9A} and~\ref{Nov3BvsNov9ARicci} which use the new
boundary conditions: first, there is an unstable exponential growth in
the highest resolution runs starting at about $t\approx 200M$, and
second, there is some complicated structure starting at about $t\approx
20M$ which is reduced and finally vanishes at the highest radial
resolution.  Consider first the unstable exponential growth.
Figure~\ref{ConstraintVsOuterBc} shows this exponential growth more
clearly by displaying the constraint violations for a range
$r_\mathrm{max}$.  The simulations in Fig.~\ref{ConstraintVsOuterBc}
have $N_r=51$ and $\ell_\mathrm{max}=11$ in each subdomain, while
$r_\mathrm{max}$ is varied by adjusting the number of subdomains.
However, in order to resolve the small incoming waves reflected off
the outer boundary, we need higher resolution near the outer boundary.
The subdomains are made narrower therefore for larger $r$, as
described below.  Figure~\ref{ConstraintVsOuterBc} shows that the
exponential growth rate of this constraint violating instability
decreases as $r_\mathrm{max}$ increases.  The growth rate of this
constraint-violating instability is about $1.7$ times larger than the
growth rate seen in the unperturbed black-hole evolutions of
Fig.~\ref{UnpertFreezingOB}, for each $r_\mathrm{max}$.  The similarity
in size and dependence on the location of the outer boundary suggests
that the instabilities in Figs.~\ref{UnpertFreezingOB} and
\ref{ConstraintVsOuterBc} may have the same basic cause.  If so then
this cause is probably bulk generated constraint violations, since the
evolutions in Fig.~\ref{ConstraintVsOuterBc} use constraint preserving
boundary conditions.
%
%--------------------BEGIN--FIGURE-----------------------------------------
%---------------------END--FIGURE-----------------------------------------

The second interesting feature seen in Figs.~\ref{Nov3BvsNov9A}
and~\ref{Nov3BvsNov9ARicci} is the complicated structure of the curves
starting at about $t\approx 20M$, when the gravitational wave passes
though the outer boundary.  This structure seems to be caused by the
(very weak) reflection of the gravitational wave pulse back into very
short wavelength ingoing waves.  Consider an outgoing wave with
coordinate velocity near one (the speed of light) as it approaches the
outer boundary.  Since the non-linear evolution equations and our
boundary conditions couple the various characteristic fields, this
wave will be mixed with and partially reflected as an ingoing wave
which propagates at the much smaller shift speed $N^i n_i\ll 1$, even
in the continuum limit.  If the original outgoing wave has wavelength
$\lambda$, then the reflected incoming wave that propagates at the
shift speed will have the much smaller wavelength $\lambda N^i n_i\ll
\lambda$.  As the outer boundary radius is increased, the amplitude of
these reflected waves is decreased.  But their wavelength also
decreases (because the shift $N^i$ decreases as $r$ increases) and
therefore these reflected waves become more difficult to resolve
numerically for larger $r_\mathrm{max}$.

One approach is to ignore these very small amplitude non-physical
reflected waves and not even attempt to resolve them.  However, these
reflected waves contribute (slightly) to the constraint quantities, so
leaving them unresolved would introduce constraint violations that are
roughly the size of the reflected waves.  If this constraint violation
is smaller than numerical truncation error in the remainder of the
domain, then no harm is done by ignoring the reflected waves.
However, for the pseudospectral simulations presented here, the
truncation error is so small elsewhere that the contributions of the
reflected waves can dominate, obscuring our convergence tests.
Therefore, we choose to resolve the reflected waves in the tests
presented here.

Another approach would be to attempt to eliminate the problem
completely by making a smarter choice for the shift $N^i$.  The radial
component of the shift could be made to approach a constant value
rather than falling to zero as $r$ increases.  This would limit the
amount of blue-shift the reflected waves could experience.
Alternatively the radial component of the shift could be made to pass
through zero and switch sign~\cite{Bardeen2002},
thus eliminating the
shift-speed incoming waves completely.  These changes would either limit
or entirely eliminate the problem, but possibly at the expense of
introducing other gauge-related difficulties. Since the choice of
gauge is not the main subject of this paper, we decided to deal
with the reflected-wave problem here by increasing the resolution in the
subdomains near the outer boundary.  For instance, the
$r_\mathrm{max}=26.9M$ curve in Fig.~\ref{ConstraintVsOuterBc} was
produced using eight subdomains with boundaries at $r=1.9M$, $6.9M$,
$11.9M$, $16.9M$, $19.4M$, $21.9M$, $23.567M$, $25.233M$, and $26.9M$,
each using $N_r=51$ collocation points.

%%%%%%%%%%%%%%%%%%%%%%%%%%%%%%%%%%%%%%%%%%%%%%%%%%%%%%%%%%%%%%%%%%%%%%%%%%%%%%%
\subsection{Angular Instability}
\label{sec:angular-stability}

In addition to the constraint violating instability and the gauge mode
instability discussed in Secs.~\ref{s:Schwarzschild} and
\ref{sec:pert-schw-black}, there are also signs of a very weak
instability that primarily affects the highest angular modes
of the evolved fields.  This angular instability has a
growth rate that increases with increasing $\ell_\mathrm{max}$, and so
this instability appears to be non-convergent.  This
instability is not evident in any of the figures shown so far, and is
negligible on the time scales of interest here except for simulations
using very large $\ell_\mathrm{max}$ and very small $r_\mathrm{max}$.
In order to see this instability clearly, we look at a quantity that
is linear in the dynamical fields and vanishes unless the instability is
present.  To this end we define
\begin{equation}
\delta\bar{K}\equiv \bar{g}^{ij} (K_{ij}-\bar K_{ij}),
\end{equation}
where $\bar{g}^{ij}$ and $\bar K_{ij}$ are exact solutions for the
three-metric and extrinsic curvature, and $K_{ij}$ is the numerical
extrinsic curvature from our simulations.  To see the angular
instability we project $\delta \bar K$ onto the spherical harmonic
basis at each radial collocation point $r_p$,
\begin{equation}
\delta \bar{K}_{p \ell m} = \int Y_{\ell m}(\theta,\varphi)
\delta\bar{K}(r_p,\theta,\varphi) \sin\theta d\theta d\varphi,
\end{equation}
and then average these over $r_p$ and $m$ by forming
\begin{equation}
\left(\delta\bar{K}^\mathrm{RMS}_{\ell}\right)^2 = N_r^{-1}(2\ell+1)^{-1}
\sum_{p,\, |m|\leq \ell} 
(\delta\bar{K}_{p \ell m})^2.
\label{eq:DeltaKrms}
\end{equation}
%
%
%--------------------BEGIN--FIGURE-----------------------------------------
\begin{figure} 
\begin{center}
\includegraphics[width=3in]{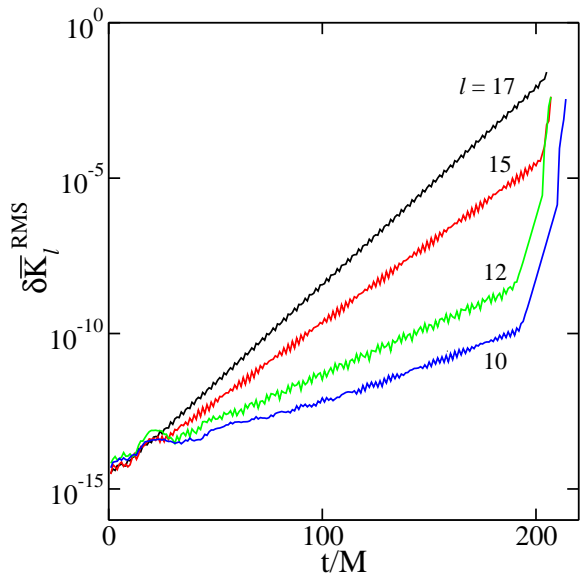}
\end{center}
\caption{Highest unfiltered components of
$\delta\bar{K}^\mathrm{RMS}_{\ell}$, defined in Eq.~(\ref{eq:DeltaKrms}),
for evolutions of unperturbed
black holes.  These evolutions use a single subdomain with $N_r=26$
and $r_\mathrm{max}=6.9M$.
\label{FanVsL-R7}}
\end{figure}
%---------------------END--FIGURE-----------------------------------------

Figure~\ref{FanVsL-R7} shows the highest unfiltered components of
$\delta\bar{K}^\mathrm{RMS}_\ell$ for evolutions of unperturbed black
holes.  These evolutions use our new boundary conditions, and are
performed on a single computational subdomain with $N_r=26$ and
$r_\mathrm{max}=6.9M$.  Each curve in Fig.~\ref{FanVsL-R7} is
generated from a run with a different $\ell_\mathrm{max}$, and the
$\delta\bar K^\mathrm{RMS}_\ell$ plotted is the one with
$\ell=\ell_\mathrm{max}-4$, the largest $\ell$ that is untouched by
our angular filtering procedure.  Each $\ell$ component grows
exponentially at a rate that increases with $\ell$.  (The extremely
rapid growth in the $\ell=10$, $12$, and $15$ components that occurs
just before the simulation crashes is presumably due to nonlinear
coupling that becomes important only when the gauge mode becomes
large.)  For all cases plotted in Fig.~\ref{FanVsL-R7} except
$\ell_{\mathrm max}=21$, the simulation crashes because of the
(convergent) gauge mode instability described earlier. For these cases
the angular instability is orders of magnitude smaller than the
unstable gauge mode for the entire duration of the simulation.  Only
for $\ell_\mathrm{max}=21$ does the angular instability dominate
before the end of the simulation.
%
%--------------------BEGIN--FIGURE-----------------------------------------
\begin{figure} 
\begin{center}
\includegraphics[width=3in]{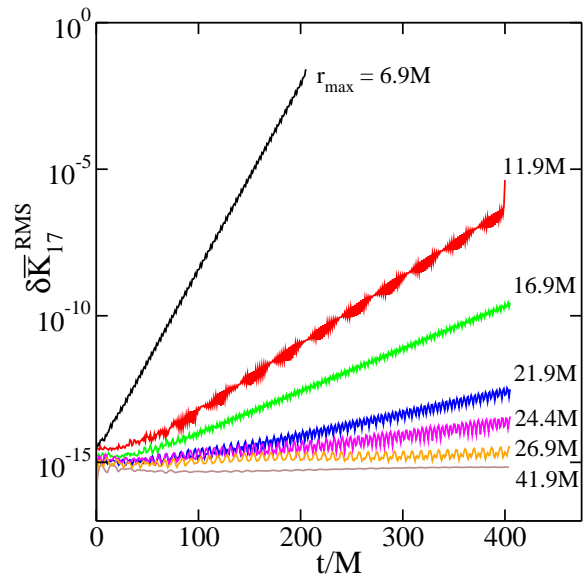}
\end{center}
\caption{$\delta\bar{K}^\mathrm{RMS}_{17}$ for unperturbed black-hole
evolutions on domains with different $r_\mathrm{max}$.  This quantity
is evaluated on the outermost domain which has width $5M$, $N_r=26$, and
$\ell_\mathrm{max}=21$.  These evolutions use our new boundary
conditions.
\label{FanVsOuterBdry}}
\end{figure}
%---------------------END--FIGURE-----------------------------------------

We have explored the behavior of this angular instability in two ways.
First we verified that the growth rate for a given $\ell$-component of
$\delta\bar K^\mathrm{RMS}_\ell$ is independent of the
$\ell_\mathrm{max}$ used to compute it.  We did this by comparing the
curves in Fig.~\ref{FanVsL-R7} with graphs of the same quantities
computed from a single run with $\ell_\mathrm{max}=21$.  The resulting
plot looks almost identical to Fig.~\ref{FanVsL-R7}, except at very
late times when the simulations begin to crash.  And second, we
explored the growth rates of the angular instability for fixed $\ell$
as a function of $r_\mathrm{max}$.  Figure~\ref{FanVsOuterBdry} shows
$\delta\bar{K}_{17}^\mathrm{RMS}$ for runs with different
$r_\mathrm{max}$.  We increased $r_\mathrm{max}$ in these runs by
adding more subdomains of width $5M$, each having $N_r=26$ and
$\ell_{\max}=21$.  Because the angular instability is largest near the
outer boundary, we compute $\delta\bar{K}_{17}^\mathrm{RMS}$ only in
the outermost subdomains for these plots.  

\begin{figure} 
\begin{center}
\includegraphics[width=3in]{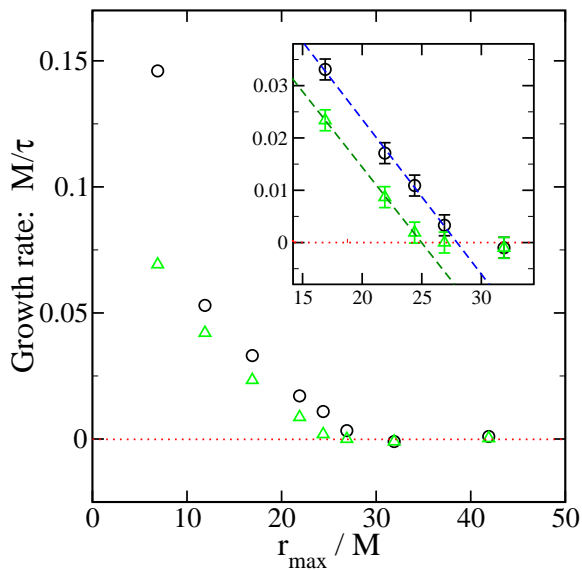}
\end{center}
\caption{Exponential growth rate of
         $\delta\bar{K}^\mathrm{RMS}_{17}$ (circles) and
         $\delta\bar{K}^\mathrm{RMS}_{12}$ (triangles) for different
         $r_\mathrm{max}$ in evolutions of unperturbed black holes with
         our new boundary conditions.  The inset shows an enlargement
         with the best linear fits through the points $16.9M\le r_{\rm max}\le
         26.9M$.
\label{FanRateVsOuterBdry}}
\end{figure}

The curves in
Fig.~\ref{FanVsOuterBdry} show that the angular instability becomes
weaker and weaker as we increase the size of the computational domain.
To study this behavior quantitatively, we plot the exponential growth
rates of $\delta\bar{K}_{17}^\mathrm{RMS}$ as a function of
$r_\mathrm{max}$; these are shown as circles in
Fig.~\ref{FanRateVsOuterBdry}.  The error bars in the inset correspond
to the variation in slopes that we extract from data segments of
different length, but these error bars are not shown in the main plot in
Fig.~\ref{FanRateVsOuterBdry} because they are the same size as the
plot symbols.  Rather than falling off gradually like
$1/r_\mathrm{max}$ as might have been expected, the growth rate in
Fig.~\ref{FanRateVsOuterBdry} appears to go to zero at a finite value
of $r_\mathrm{max}$.  This finite value of $r_\mathrm{max}$ depends on
$\ell$; to show this dependence, we also plot in
Figure~\ref{FanRateVsOuterBdry} the exponential growth rates of
$\delta\bar{K}_{12}^\mathrm{RMS}$ (shown as triangles in the plot) for
the same runs.  The best linear fits through the $\ell=17$ and
$\ell=12$ growth rates have the same slope, but the growth rate of the
$\ell=12$ mode goes to zero at a smaller $r_\mathrm{max}$.
Figure~\ref{FanRateVsOuterBdry} suggests that for a given
$\ell_{\max}$ there are no angular instabilities at all when
$r_\mathrm{max}$ is sufficiently large.  No angular instability has
been detected in any of our runs with $\ell_\mathrm{max}\leq 21$,
$t\leq300M$, and $r_\mathrm{max}>27M$.  Our results also suggest that
for a given $r_\mathrm{max}$, an angular instability with arbitrarily
large growth rate could be found by making $\ell_\mathrm{max}$
sufficiently large.
%
%--------------------BEGIN--FIGURE-----------------------------------------
%---------------------END--FIGURE-----------------------------------------

We see no angular instability at all when we use freezing boundary
conditions.  Furthermore, the angular instability shown in
Figs.~\ref{FanVsL-R7}--\ref{FanRateVsOuterBdry} is present whether or not
we use the physical boundary condition, Eq.~(\ref{eq:physicalBc}), or
the gauge boundary conditions, Eqs.~(\ref{e:freezez4}) and
(\ref{e:kfreezebc}).  [Although as noted before, an angular
instability dominates when Eq.~(\ref{e:kfreezebc}) is imposed on all
the spherical harmonic components of $U^{1-}$.]  It is unclear whether
this angular instability is due to our numerical method or whether the
new constraint-preserving boundary conditions yield an ill-posed
initial-boundary-value problem at the continuum level.  
We note however that for the
resolutions and time-scales of interest here, this instability remains
small and can be controlled quite effectively by modestly increasing
the radius of the outer boundary.  And the numerical evolutions, such
as those in Figs.~\ref{Nov3BvsNov9A}--\ref{ConstraintVsOuterBc},
produced by these methods do appear to be accurate solutions of the
Einstein equations: both the constraints $||\kernedcalC||/||\partial u||$ and
the four-dimensional Ricci tensor $||R||/||R^\mathrm{RMS}||$ can be
made arbitrarily small.  Therefore from a practical point of view it
may not matter whether these boundary conditions are formally
well-posed, or that our computational methods contain a very mild
non-convergent angular instability.

\section{Discussion}
\label{s:Discussion}

This paper constructs new boundary conditions for the KST form of the
Einstein evolution system that are designed to prevent the influx of
constraint violations and physical gravitational waves into the
computational domain.  From a mathematical point of view, these
boundary conditions are Neumann-like (in the sense that they place
conditions on the normal derivatives of the incoming characteristic
fields).  Boundary conditions of this type have not been studied as
comprehensively as the simpler ``maximally-dissipative'' boundary
conditions (which are Dirichlet-like in that they place
conditions on the values of incoming characteristic fields
themselves).  Rigorous mathematical well-posedness theorems do not yet
exist for hyperbolic evolution systems with these Neumann-like
boundary conditions.  So additional mathematical analysis of these
conditions is urgently needed to determine whether they are well-posed
both at the continuum differential equation level and the discrete
numerical level.  Should this analysis reveal that these conditions
are ill-posed, then alternate ways of preventing the influx of
constraint violations and physical gravitational waves in these
systems would be needed even more urgently.

Our numerical tests of these new constraint preserving and physical
boundary conditions show them to be quite effective: constraint violations
can be reduced to roundoff-level errors in dynamical black-hole evolutions. 
Nevertheless,
several weak instabilities did appear in our numerical results, and
additional work is needed to sort out exactly what their causes are
and what methods can be used to control them.  Are the constraint violating
instabilities seen in the evolutions of Figs.~\ref{UnpertFreezingOB}
and \ref{ConstraintVsOuterBc} really caused by bulk constraint
violating terms in the constraint evolution equations (as we surmise),
and can optimal constraint projection methods such as those developed
for the scalar field system~\cite{Holst2004} be used to control them?
Can the gauge instabilities seen in Fig.~\ref{Oct21DError} be
controlled by the introduction of dynamical evolution equations for
the lapse and shift?  Are the non-convergent angular instabilities
seen in Figs.~\ref{FanVsL-R7} and \ref{FanVsOuterBdry} a symptom of
ill-posedness of these boundary conditions at the continuum or the
discrete numerical level?  And what can be done to cure these problems?

\acknowledgments We thank Olivier Sarbach, Saul Teukolsky, and Manuel
Tiglio for helpful discussions concerning this work. Some of the
computations for this project were performed with the Tungsten cluster
at NCSA.  This work was supported in part by NSF grants PHY-0099568,
PHY-0244906 and NASA grants NAG5-10707, NAG5-12834 at Caltech, and NSF
grants PHY-0312072, PHY-0354631 at Cornell. LTB was supported by
a National Research Council Research Associateship Award at the
Jet Propulsion Laboratory.

%%%%%%%%%%%%%%%%%%%%%%%%%%%%%%%%%%%%%%%%%%%%%%%%%%%%%%%%%%%%%%%%%%%%%%%%%%%%%%%
\appendix
\section{KST Characteristic Fields}
\label{s:AppendixA}

Explicit expressions are given here for the characteristic fields of
the KST form of the Einstein evolution system.  These characteristic
fields consist of the collection $u^{\hat\alpha}=$ $\{g_{ij}, Z^1$, $Z^2_i$,
$Z^3_i$, $Z^4_i$, $Z^5_{ij}$, $Z^6_{kij}$, $U^{1\pm}$, $U^{2\pm}_i$,
$U^{3\pm}$, $U^{4\pm}_{ij}\}$, and can be
written in terms of the principal evolution fields,
Eq.~(\ref{e:characteristicparts}):
\begin{eqnarray}
Z^1&=&\gamma_3n^iD^1_i-2(1+\gamma_4)n^iD^2_i,\label{e:z1}\\
Z^2_i&=&\gamma_4P^j{}_iD^1_j-(\gamma_3+2\gamma_4)P^j{}_in^kn^lD_{jkl},
\label{e:z2}\\
Z^3_i&=&3P^j{}_iD^1_j-2P^j{}_iD^2_j-4P^j{}_in^kn^lD_{jkl},
\label{e:z3}\\
Z^4_i&=&+48v_2^2n^lP^j{}_in^kD_{ljk}+2\gamma_4(5-9\gamma_2)P^j{}_iD^2_j
\nonumber\\
&&\!\!\!\!\!\!\!+3(1-3\gamma_2-4\gamma_0)(4-\gamma_3)\bigl(P^j{}_iD^1_j
+P^j{}_in^kn^lD_{jkl}\bigr)
\nonumber\\
&&\!\!\!\!\!\!\!
-2(6+\gamma_4)(5-9\gamma_2) P^j{}_in^kn^lD_{jkl},\\
Z^5_{ij}&=&\bigl(P^a{}_iP^b{}_j-\half P_{ij}P^{ab}\bigr)n^kD_{abk},
\label{e:z5}\\
Z^6_{kij}&=&P^{cab}_{kij}D_{cab},\label{e:z6}\\
U^{1\pm}&=&\pm\bigl[1+2v_1^2+(1+2\gamma_1)q\bigr]n^iD^1_i
\nonumber\\
&&\!\!\!\!\!\!
 - v_1(1-q)P^{ij}K_{ij}  
+2v_1n^in^j \bigl[K_{ij} \pm v_1 n^kD_{kij}\bigr]\nonumber\\
&&\!\!\!\!\!\!
\mp\bigl[1-3\gamma_2+(1+2\gamma_1+\gamma_2)q\bigr] n^iD^2_i,
\label{e:U1}\\
U^{2\pm}_i&=&\pm 2v_2 n^k P^j{}_i K_{jk}
+(1+2\gamma_0)P^j{}_iD^1_j,
\nonumber\\
&&\!\!\!\!\!\!\!
-(1-\gamma_2)P^j{}_iD^2_j+(2\gamma_0-\gamma_2)P^j{}_in^kn^lD_{jkl},\\
U^{3\pm}&=&\pm(1+2\gamma_1)n^iD^1_i \mp(1+2\gamma_1+\gamma_2)n^iD^2_i
\nonumber\\
&&\!\!\!\!\!\!\!
+ v_3 P^{ij}K_{ij},\\
U^{4\pm}_{ij}&=&\bigl(P^a{}_iP^b{}_j-\half P_{ij}P^{ab}\bigr)\nonumber\\
&&\!\!\!\!\!\!\!\times\Bigl[K_{ab} \pm n^kD_{kab} 
  \mp (1+\gamma_2)n^kD_{(ab)k}\Bigr].
\label{e:u4}
\end{eqnarray}
In these expressions the two distinct traces of $D_{kij}$ are
written as
\begin{eqnarray}
D^1_{i}&=&P^{jk}D_{ijk},\\
D^2_{i}&=&P^{jk}D_{kij},
\end{eqnarray}
where 
\begin{equation}
P^{ij}=g^{ij}-n^in^j,
\end{equation}
is the projection orthogonal to $n_i$.  The quantities $v_1$, $v_2$
and $v_3$ are defined in Eqs.~(\ref{e:v1})--(\ref{e:v3}), and $q$ is
given by
\begin{eqnarray}
q&=&\frac{1+3v_1^2-4v_2^2}{v_1^2-v_3^2}.
\label{eq:CharFieldQdefinition}
\end{eqnarray}
Finally, the projection tensor $P^{cab}_{kij}$ is defined by
\begin{eqnarray}
P^{cab}_{kij}&=&P^c{}_kP^{(a}{}_iP^{b)}{}_j-\threefourths P_{ij}P^c{}_kP^{ab}
+\half P^{c(a}P^{b)}{}_kP_{ij}\nonumber\\
&&+\half P^{ab}P^{c}{}_{(i}P_{j)k}
-P^{c(a}P^{b)}{}_{(i}P_{j)k}.\label{e:projectiontensor}
\end{eqnarray}
These expressions are the completely general forms of the
characteristic fields for the KST system.  They reduce to the
expressions for the restricted case $v_1^2=v_2^2=v_3^3=1$ published
previously in Ref.~\cite{Kidder2001}.

The characteristic fields $U^{1\pm}$, $U^{2\pm}_i$ and $U^{3\pm}$ have
characteristic speeds $\pm v_1$, $\pm v_2$, and $\pm v_3$ respectively
(relative to the hypersurface normal observers); the fields $U^{4\pm}_{ij}$
have speeds $\pm 1$; the fields $\{Z^1$, $Z^2_i$, $Z^3_i$, $Z^4_i$,
$Z^5_{ij}$, $Z^6_{kij}\}$ all have characteristic speed zero.  
The characteristic fields are linearly independent (so
the KST evolution system is strongly hyperbolic) if 
$v_1\neq0$, $v_2\neq 0$, $v_3\neq 0$, and $v_1\neq v_3$.  
In the case when $v_1=v_3$, the characteristic fields
$U^{1\pm}$ given in Eq.~(\ref{e:U1}) are not defined because the
quantity $q$ given in Eq.~(\ref{eq:CharFieldQdefinition}) is not
defined.  We find that there are nevertheless a complete set of
characteristic fields in the $v_1=v_3\neq0$ case so long as
\begin{equation}
1+3v_1^2=4v_2^2.\label{e:speedcondition}
\end{equation}  
In this case, the characteristic fields $U^{1\pm}$ are just given by
the expression in Eq.~(\ref{e:U1}) with $q=0$.  Any other constant
value of $q$ is also acceptable, with the result being a redefinition
of $U^{1\pm}$ by the addition of $q$ times $U^{3\pm}$ (which has the
same characteristic speed in this case).  The choice $q=0$ is probably
the simplest, but other choices might be desirable under some
circumstances.  For example, for
symmetric hyperbolic systems it might be better
to make the eigenvectors mutually orthogonal in terms of the
symmetrizer metric.
The symmetric hyperbolicity of the KST system is discussed
in Appendix B of Ref.~\cite{Lindblom2003}.

It is also useful to have explicit expressions for the inverse
transformation, $u^\alpha=e^\alpha{}_{\hat\beta}u^{\hat\beta}$, that
expresses the principal evolution fields in terms of the
characteristic fields.  These inverse transformations for the general
KST form of the Einstein evolution system are given by
\begin{eqnarray}
K_{ij}&=&n_in_j\biggl[\frac{1-q}{4v_3}(U^{3+}+U^{3-})
+\frac{U^{1+}+U^{1-}}{4v_1}\biggr]
\nonumber\\
&&\!\!\!\!\!
+\frac{n{}_{(i}U^{2+}_{j)}-n{}_{(i}U^{2-}_{j)}}{2v_2}
+P_{ij}\frac{U^{3+}+U^{3-}}{4v_3}\nonumber\\
&&\!\!\!\!\!
+\half (U^{4+}_{ij}+U^{4-}_{ij}),
\label{e:Kij}\\
D_{kij}&=&n_kn_in_jn^cn^an^bD_{cab}+n_in_jP^c{}_k n^an^bD_{cab}\nonumber\\
&&\!\!\!\!\!+2n_kn_{(i}P^a{}_{j)}n^bn^cD_{cab}
+\half n_k(U^{4+}_{ij}-U^{4-}_{ij})\nonumber\\
&&\!\!\!\!\!+(1+\gamma_2)n_kZ^5_{(ij)}+2Z^5_{k(i}n_{j)}+Z^6_{kij}
\nonumber\\
&&\!\!\!\!\!+\half n_k P_{ij}n^cD^1_c+\threefourths P_{ij}P^c{}_kD^1_c
-\half P_{k(i}P^c{}_{j)}D^1_c\nonumber\\
&&\!\!\!\!\!+n_{(i}P_{j)k}n^cD^2_c-\half P_{ij}P^c{}_kD^2_c
+P_{k(i}P^c{}_{j)}D^2_c.\qquad\label{e:Dkij}\nonumber\\
\end{eqnarray}
The terms involving $D_{kij}$ on the right sides of Eq.~(\ref{e:Dkij})
are given by
\begin{eqnarray}
&&\!\!\!\!\!\!
n^cn^an^bD_{cab}=\bigl[v_1^2(1+2\gamma_1+\gamma_2)
+\gamma_2(2+3\gamma_1)\bigr]\frac{Z^1}{2v_1^2v_3^2}\nonumber\\
&&
+
\frac{1+\gamma_3+\gamma_4-q}{4v_3^2}
(U^{3+}-U^{3-})
+\frac{U^{1+}-U^{1-}}{4v_1^2},\\
&&\!\!\!\!\!\!
P^c{}_in^an^bD_{cab}=-\frac{1-3\gamma_2-4\gamma_0}{8v_2^2}Z^2_i
+\frac{\gamma_4(1-\gamma_2)}{8v_2^2}Z^3_i\nonumber\\
&&-\frac{\gamma_4}{8v_2^2}(U^{2+}_i+U^{2-}_i),\\
&&\!\!\!\!\!\!
P^a{}_in^cn^bD_{cab}=-\frac{5-9\gamma_2}{16v_2^2}Z^2_i
+\frac{4-\gamma_3}{16v_2^2}(U^{2+}_i+U^{2-}_i)
\nonumber\\
&&-\bigl[3(1-\gamma_2)(4-\gamma_3)-\gamma_4(5-9\gamma_2)\bigr]
\frac{Z^3_i}{48v_2^2}+\frac{Z^4_i}{48v_2^2},\nonumber\\
\\
&&\!\!\!\!\!\!
n^cD^1_c=\frac{1+\gamma_4}{2v_3^2}(U^{3+}-U^{3-})
-(1+2\gamma_1+\gamma_2)\frac{Z^1}{2v_3^2},\nonumber\\
\\
&&\!\!\!\!\!\!
n^cD^2_c=\frac{\gamma_3}{4v_3^2}(U^{3+}-U^{3-})
-(1+2\gamma_1)\frac{Z^1}{2v_3^2},\\
&&\!\!\!\!\!\!
P^c{}_iD^1_c=(1-\gamma_2)(\gamma_3+2\gamma_4)\frac{Z^3_i}{8v_2^2}
-(2-3\gamma_2+2\gamma_0)\frac{Z^2_i}{4v_2^2}\nonumber\\
&&
-\frac{\gamma_3+2\gamma_4}{8v_2^2}(U^{2+}_i+U^{2-}_i),\\
&&\!\!\!\!\!\!
P^c{}_iD_c^2=\bigl[\gamma_3(1+2\gamma_0)+\gamma_4(2-\gamma_2+6\gamma_0)\bigr]
\frac{Z^3_i}{8v_2^2}
\nonumber\\
&&-(3\gamma_3+2\gamma_4)\frac{U^{2+}_i+U^{2-}_i}{16v_2^2}
-(4-3\gamma_2+14\gamma_0)\frac{Z^2_i}{8v_2^2}.\nonumber\\
\end{eqnarray}
%

%%%%%%%%%%%%%%%%%%%%%%%%%%%%%%%%%%%%%%%%%%%%%%%%%%%%%%%%%%%%%%%%%%%%%%%%%%%%%%%
\section{Hyperbolicity of Constraint Evolution}
\label{s:AppendixB}

In this Appendix we evaluate the hyperbolicity of the KST constraint
evolution system, Eq.~(\ref{e:hamdot})--(\ref{e:c4dot}).  This system
can be written in the more abstract form
\begin{eqnarray}
\partial_t c^A + A^{kA}{}_B \partial_k c^B = F^A{}_B c^B.
\end{eqnarray}
In Sec.~\ref{s:ConstraintEvolutionSystem} 
we demonstrated the strong hyperbolicity of the KST
constraint evolution evolution system by
showing that the characteristic matrices $n_kA^{kA}{}_B$ have a
complete set of eigenvectors.  Here we determine the conditions under
which this constraint evolution system is also symmetric hyperbolic,
{\it i.e.,} that there exists a symmetric positive-definite $S_{AB}$
(the symmetrizer) with the property that it symmetrizes the
characteristic matrices: $A^k_{AB}\equiv S_{AC}A^{kC}{}_B = A^k_{BA}$.

The most general symmetrizer for the constraint evolution
system can be written conveniently by defining the following
quantities associated with the constraint ${\cal C}_{klij}$:
\begin{equation}
{\cal D}_{kij} = \epsilon_{k}{}^{ab}{\cal C}_{abij},
\end{equation}
where $\epsilon_{kij}$ is the spatial volume element.  We also define
\begin{eqnarray}
{\cal D}^1_i &=& g^{jk}{\cal D}_{ijk},\\ 
{\cal D}^2_i &=& g^{jk}{\cal D}_{jik},\\
\tilde {\cal D}_{kij} &=& {\cal D}_{kij} + \fifth \bigl[
{\cal D}^1_{(i} g_{j)k} - 2{\cal D}^1_{k} g_{ij}\nonumber\\
&&\qquad\qquad+ {\cal D}^2_{k} g_{ij} - 3{\cal D}^2_{(i} g_{j)k}\bigr].
\end{eqnarray}
Then the most general symmetrizer constructed from the metric $g_{ij}$
is:
\begin{eqnarray}
&&\!\!\!\!\!\!\!\!\!\!
dS^2 =S_{AB} dc^A dc^B =\nonumber\\
&&\!\!\!\chi_1 d{\cal C}^2 + \chi_2 g^{ia} d{\cal C}_id{\cal C}_a
+\chi_3 g^{ia}g^{jb}g^{kc} d\tilde{\cal D}_{(kij)}d\tilde{\cal D}_{(cab)}
\nonumber\\
&&\!\!\!+\chi_4 g^{ia}g^{jb}g^{kc} 
\bigl[d\tilde{\cal D}_{kij}-d\tilde{\cal D}_{(kij)}\bigr] 
\bigl[d\tilde{\cal D}_{cab}-d\tilde{\cal D}_{(cab)}\bigr]\nonumber\\
&&\!\!\!+\chi_5 g^{ia} d{\cal D}^1_id{\cal D}^1_a
+\chi_6 g^{ia} d{\cal D}^2_id{\cal D}^2_a
+2\chi_7 g^{ia} d{\cal D}^1_id{\cal D}^2_a.\nonumber\\
\end{eqnarray}
The conditions needed to ensure that this symmetrizer is positive
definite are $\chi_a>0$ for $a=1,\ldots,6$ and
$\chi_5\chi_6>\chi_7^2$.  There are no cross terms
between the momentum constraint ${\cal C}_i$ and the ${\cal D}^1_i$ or
${\cal D}^2_i$ constraints, because the resulting terms would have the
wrong parity.  Using the principal parts of the KST constraint
evolution system given in Eqs.~(\ref{e:hamdot})--(\ref{e:c4dot}), we
find that the conditions needed for symmetric hyperbolicity are
\begin{eqnarray}
0 &=& \chi_1(1 - \frac{1}{2}\gamma_3 + \gamma_4) 
- \chi_2(1 + 2\gamma_1),\label{e:symm1} \\
0 &=& 2 \chi_5(\gamma_3 + 3 \gamma_4) + \chi_2\gamma_0 
+ \chi_7 (2\gamma_4-\gamma_3),\\
0 &=& \frac{1}{2} \chi_2 
- \chi_6(\gamma_3 - 2\gamma_4)
+ \chi_7(2\gamma_3 + 6\gamma_4), \\
0 &=& \frac{1}{2} \chi_2 \gamma_2 + \chi_4 \gamma_3.\label{e:symm4}
\end{eqnarray}

The problem now is to determine when these symmetrization conditions
can be satisfied.  For simplicity we focus attention here on the case
where all of the characteristic speeds of the principal evolution
system have the physical values: $0$, $\pm 1$.  These conditions on
the characteristic speeds implies the following conditions on the
parameters:
\begin{eqnarray}
\gamma_0&=&\half,\label{e:physv1}\\
\gamma_3&=&\frac{-8}{4\gamma_2+(5+3\gamma_2)(1+2\gamma_1)},\label{e:physv2}\\
\gamma_4&=&\frac{1-\gamma_2-(1+2\gamma_1)(5+3\gamma_2)}
{4\gamma_2+(5+3\gamma_2)(1+2\gamma_1)}.\label{e:physv3}
\end{eqnarray}
The analysis in Lindblom and Scheel~\cite{Lindblom2002} shows that the
principal evolution system is symmetric hyperbolic in this case for
all values of the parameters $\gamma_1$ and $\gamma_2$ that satisfy
the following inequalities:
\begin{eqnarray}
&&-\fivethirds<\gamma_2<0,\label{e:prinsh1}\\
&&4\gamma_2+(5+3\gamma_2)(1+2\gamma_1)\neq 0,\label{e:prinsh2}
\end{eqnarray}
Substituting the conditions Eqs.~(\ref{e:physv1})--(\ref{e:physv3})
into the symmetry conditions, Eqs.~(\ref{e:symm1})--(\ref{e:symm4}),
we find the following conditions on the $\chi_a$:
\begin{eqnarray}
\chi_2&=&\frac{\chi_1(5+3\gamma_2)}
{(1+2\gamma_1)[4\gamma_2+(5+3\gamma_2)(1+2\gamma_1)]},
\label{e:symmchi2}\\
\chi_4&=&\frac{\chi_1\gamma_2(5+3\gamma_2)}{16(1+2\gamma_1)},
\label{e:symmchi4}\\
\chi_5&=&\frac{\chi_1(5+3\gamma_2)
-8\chi_7(1+2\gamma_1)[2\gamma_2+\gamma_1(5+3\gamma_2)]}
{8(5+3\gamma_2)(2+7\gamma_1+6\gamma_1^2)},\nonumber\\
&&\label{e:symmchi5}\\
\chi_6&=&\frac{(5+3\gamma_2)[\chi_1
-8\chi_7(2+7\gamma_1+6\gamma_1^2)]}
{8(1+2\gamma_1)[2\gamma_2+\gamma_1(5+3\gamma_2)]}.
\label{e:symmchi6}
\end{eqnarray}
The constraint evolution system is symmetric hyperbolic iff $\chi_a>0$
for $a=\{1,2,3,4,5,6\}$ and $\chi_5\chi_6>\chi_7^2$.  If the principal
evolution system is symmetric hyperbolic then $5+3\gamma_2>0$ [from
Eq.~(\ref{e:prinsh1})], thus we see from Eq.~(\ref{e:symmchi4}) that
we must have 
\begin{eqnarray}
1+2\gamma_1<0,\label{e:limgm1}
\end{eqnarray}
or equivalently $\gamma_1<-\half$ if the constraint evolution system
is to be symmetric hyperbolic as well.  This condition guarantees
that $4\gamma_2+(5+3\gamma_2)(1+2\gamma_1)<0$ and so the second
inequality in Eq.~(\ref{e:prinsh2}) is automatically satisfied
in this case.  Consequently Eq.~(\ref{e:symmchi2}) places no further
restrictions on the parameters $\gamma_1$ and $\gamma_2$.  

To analyze the inequalities on $\chi_5$ and $\chi_6$ implied by
Eqs.~(\ref{e:symmchi5}) and (\ref{e:symmchi6}), 
we restrict our attention to the simple case where
$\chi_7$ can be set to zero.  In this case Eqs.~(\ref{e:symmchi5}) and
(\ref{e:symmchi6}) simplify considerably:

\begin{eqnarray}
\chi_5&=&\frac{\chi_1}
{8(2\gamma_1+1)(3\gamma_1+2)},\label{e:symmchi5simp}\\
\chi_6&=&\frac{(5+3\gamma_2)\chi_1}
{8(1+2\gamma_1)[2\gamma_2+\gamma_1(5+3\gamma_2)]}.\label{e:symmchi6simp}
\end{eqnarray}
Equation~(\ref{e:symmchi6simp}) guarantees that the
coefficient $\chi_6$ is positive without any additional restrictions,
while Eq.~(\ref{e:symmchi5simp}) shows that the parameter $\gamma_1$
must be further restricted by the inequality
\begin{eqnarray}
\gamma_1<-\twothirds,
\end{eqnarray}
in order to ensure that $\chi_5$ is positive.  This argument shows
then that we can choose the parameters $\chi_a>0$ for
$a=\{1,2,3,4,5,6\}$ and $\chi_7=0$ for any values of the parameters
$\gamma_1$ and $\gamma_2$ in the ranges:  
\begin{eqnarray}
\gamma_1&<&-\twothirds,\label{e:constsymmlim1}\\
-\fivethirds&<&\gamma_2<0.\label{e:constsymmlim2}
\end{eqnarray}

We have not yet found the minimal set of restrictions on the
parameters that allows the KST constraint evolution system to be
symmetric hyperbolic for any values of the characteristic speeds, and
for any values of the parameter $\chi_7$.  We have limited our
numerical experiments so far to the regions of this parameter space
where both the principal and the constraint evolution systems are
known to be symmetric hyperbolic.

%%%%%%%%%%%%%%%%%%%%%%%%%%%%%%%%%%%%%%%%%%%%%%%%%%%%%%%%%%%%%%%%%%%%%%%%%%%%%%%
\bibstyle{prd} 
\bibliography{References}

\end{document}